\documentclass{llncs}
\usepackage{t1enc}
\usepackage[utf8]{inputenc}
 \usepackage{times}

\usepackage{amsfonts}

\newcommand{\set}[1]{\left\{ #1\right\}}

\newcommand{\sodass}{\,:\,}
\newcommand{\setGilt}[2]{\left\{ #1\sodass #2\right\}}

\newcommand{\realrange}[2]{\left[#1, #2\right]}

\newcommand{\unitrange}[2]{\realrange{0}{1}}

\newcommand{\Oh}[1]{\mathcal{O}\!\left( #1\right)}

\newcommand{\llabel}[1]{\label{\labelprefix:#1}}
\newcommand{\labelprefix}{} 

\newcommand{\discussionsize}{\small}

\newenvironment{code}{\noindent
\begin{tabbing}%
\hspace{2em}\=\hspace{2em}\=\hspace{2em}\=\hspace{2em}\=\hspace{2em}\=%
\hspace{2em}\=\hspace{2em}\=\hspace{2em}\=\hspace{2em}\=\hspace{2em}\=%
\kill}{\end{tabbing}}

\newcommand{\labelcommand}{}
\newcommand{\captiontext}{}
\newsavebox{\codeparam}
\newcounter{lineNumber}
\newenvironment{disscodepos}[3]{%
\renewcommand{\labelcommand}{#2}%
\renewcommand{\captiontext}{#3}%
\sbox{\codeparam}{\parbox{\textwidth}{#3}}%
\begin{figure}[#1]\begin{center}\begin{code}\setcounter{lineNumber}{1}}{%
\end{code}\end{center}\caption{\llabel{\labelcommand}\captiontext}\end{figure}}

{\end{disscodepos}}

\newcommand{\Is}       {:=}

\newdimen\endofsize\endofsize=0.5em
\def\endofbeweis{~\quad\hglue\hsize minus\hsize
                 \hbox{\vrule height \endofsize width
\endofsize}\par}

\usepackage{epsfig}
\usepackage{url}
\usepackage{amsmath}
\usepackage{amssymb}
\usepackage{pdflscape}
\usepackage{numprint}
\npdecimalsign{.} 
\usepackage{theorem}
\usepackage{makeidx}
\usepackage[usenames,dvipsnames]{xcolor}
\usepackage{hyperref}
\usepackage{xspace}
\usepackage{wrapfig}
\usepackage{graphics}
\usepackage{todonotes}
\usepackage[left,pagewise] {lineno}

\def\MdR{\ensuremath{\mathbb{R}}}

\newcommand{\Id}[1]{\ensuremath{\text{{\sf #1}}}}
\newcommand{\algorithmname}{MulMent}
\newcommand{\algoname}{\algorithmname}

\newcommand{\ie}{i.\,e.\xspace}
\newcommand{\eg}{e.\,g.\xspace}
\newcommand{\etal}{et~al.\xspace}

\usepackage{color}

\newcommand{\mnoe}[1]{{\color{red}[MN: #1]}}
\newcommand{\hmey}[1]{{\color{orange}[HM: #1]}}
\newcommand{\csch}[1]{{\color{blue}[CS: #1]}}
  \renewcommand{\mnoe}[1]{}
  \renewcommand{\hmey}[1]{}
  \renewcommand{\csch}[1]{}

\newcommand{\mytitle}{Drawing Large Graphs by \\ Multilevel Maxent-Stress Optimization}

\begin{document}
\title{\mytitle}
\author{Henning Meyerhenke, Martin N\"ollenburg, Christian Schulz}
\institute{ 
	Institute of Theoretical Informatics \\ 
	Karlsruhe Institute of Technology (KIT),
	Karlsruhe, Germany \\
	Email: \email{\{meyerhenke, noellenburg, christian.schulz\}@kit.edu}
}
\date{}

\maketitle
\begin{abstract}
Drawing large graphs appropriately is an important step for the visual analysis of data from
real-world networks. Here we present a novel multilevel algorithm to compute a graph layout with respect to a 
recently proposed metric that combines layout stress and entropy.
As opposed to previous work, we do not solve the linear systems of the maxent-stress metric with a typical 
numerical solver. Instead we use a simple local iterative scheme within a multilevel approach.
To accelerate local optimization, we approximate long-range forces and use shared-memory parallelism.
Our experiments validate the high potential of our approach, which is particularly appealing for dynamic graphs.
In comparison to the previously best maxent-stress optimizer, which is sequential, our parallel implementation
is on average 30 times faster already for static graphs (and still faster if executed on one thread)
while producing a comparable solution quality.
\end{abstract}

\section{Introduction}
\label{sec:intro}

Drawing large networks (or graphs, we use both terms interchangeably) with hundreds of thousands of nodes and
edges has a variety of relevant applications. One of them can be interactive visualization, which helps humans working on
graph data to gain insights about the properties of the data. If a very large high-end display is not available for such
purpose, a hierarchical approach allows the user to select an appropriate zoom level~\cite{abello-ask-06}.
Moreover, drawings of large graphs can also be used as a preprocessing step in high-performance applications~\cite{Kirmani:2013:SPG:2503210.2503280}.

One very promising class of layout algorithms in this context is based on the
\emph{stress} of a graph. Such algorithms can for instance be used for drawing graphs with fixed
distances between vertex pairs, provided \textit{a priori} in a distance matrix~\cite{gkn-gdsm-05}.
More recently, Gansner \etal~\cite{ghn-mmgl-13} proposed a similar model 
that includes besides the stress an additional entropy term (hence its name \emph{maxent-stress}). 
While still using shortest path distances, this model often results in more satisfactory layouts for large networks.
The optimization problem can be cast as solving Laplacian 
linear systems successively. Since each right-hand side in this succession 
depends on the previous solution, many linear systems need
to be solved until convergence -- more details can be found in Section~\ref{par:maxent}.

\vspace*{0.75ex}
\emph{Motivation.}
\label{par:motivation}
We want to employ this maxent-stress model for drawing large networks quickly.
Yet, solving many large Laplacian linear systems can be quite costly.
A conjugate gradient solver (used in~\cite{ghn-mmgl-13}) is easy to implement but has superlinear running time. 
Solvers with provably nearly-linear running time exist but are not yet competitive with established methods 
in practice (see~\cite{HoskeLMW15nearly} for an experimental comparison). Multigrid 
methods~\cite{DBLP:journals/cviu/KoutisMT11,DBLP:journals/siamsc/LivneB12} for Laplacian systems may seem 
appealing in this context, but their setup phase building the multigrid hierarchy can be expensive for large graphs.

Gansner \etal~\cite{ghn-mmgl-13} also suggested (but did not use) a simpler iterative refinement procedure for solving 
their optimization problem. This procedure would be slow to converge if used unmodified. However, if designed and 
implemented appropriately, it has the potential for fast convergence even on large graphs. Moreover, as already observed
in~\cite{ghn-mmgl-13}, it has high potential for parallelism and should work well on dynamic graphs by profiting from  
previous solutions.

\vspace*{0.75ex}
\emph{Outline and Contribution.}
\label{par:outline-contribution}
The main contribution of this paper is to make the alternative iterative local optimizer suggested by 
Gansner \etal~\cite{ghn-mmgl-13} (for details on this and other related work see Section~\ref{sec:prelim}) 
usable and fast in practice. 
To this end, we design and implement a multilevel algorithm tailored to large networks (see Section~\ref{sec:algo}). 
The employed coarsening algorithm for building the multilevel hierarchy can control the trade-off between
the number of hierarchy levels and convergence speed of the local optimizer.
One property of the local optimizer we exploit is its high degree of parallelism. 
Further acceleration is obtained by approximating long-range forces. 
To this end, we use coarser representatives stored in the multilevel hierarchy.

\begin{figure}[bt]
\centering
\vspace*{-.25cm}
\includegraphics[width=3.25cm]{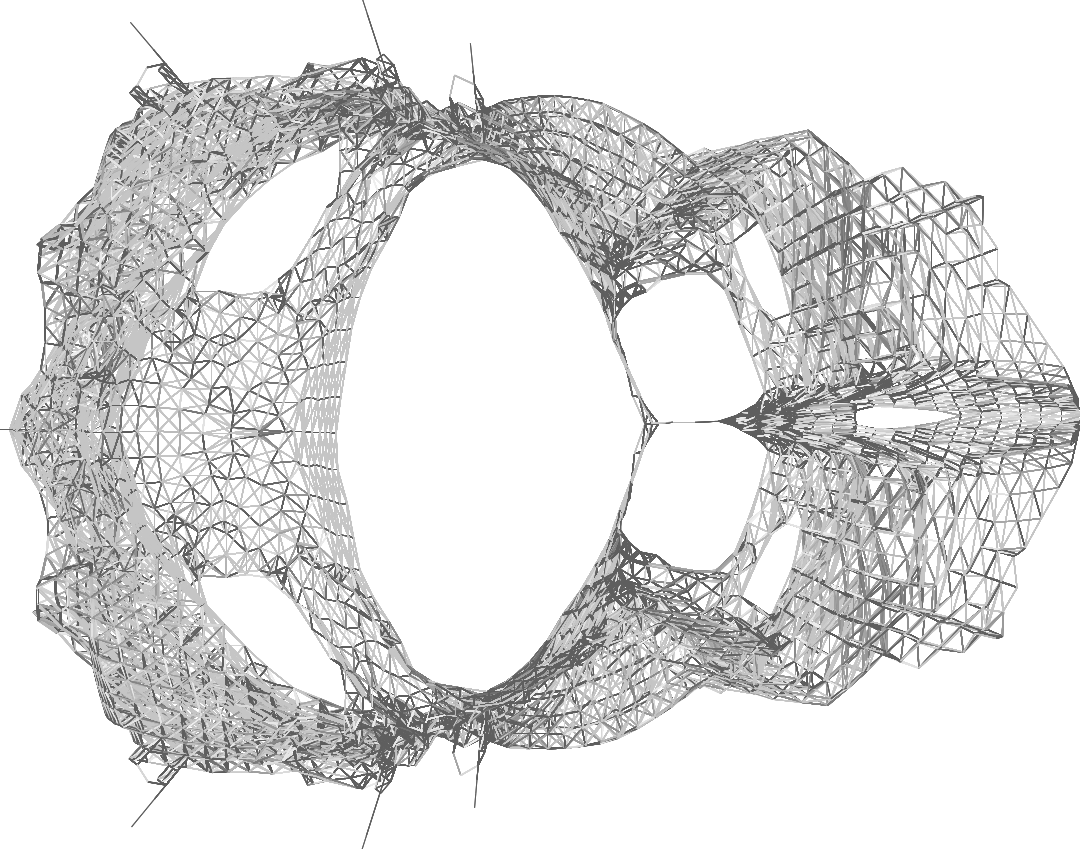} 
\includegraphics[width=3.75cm]{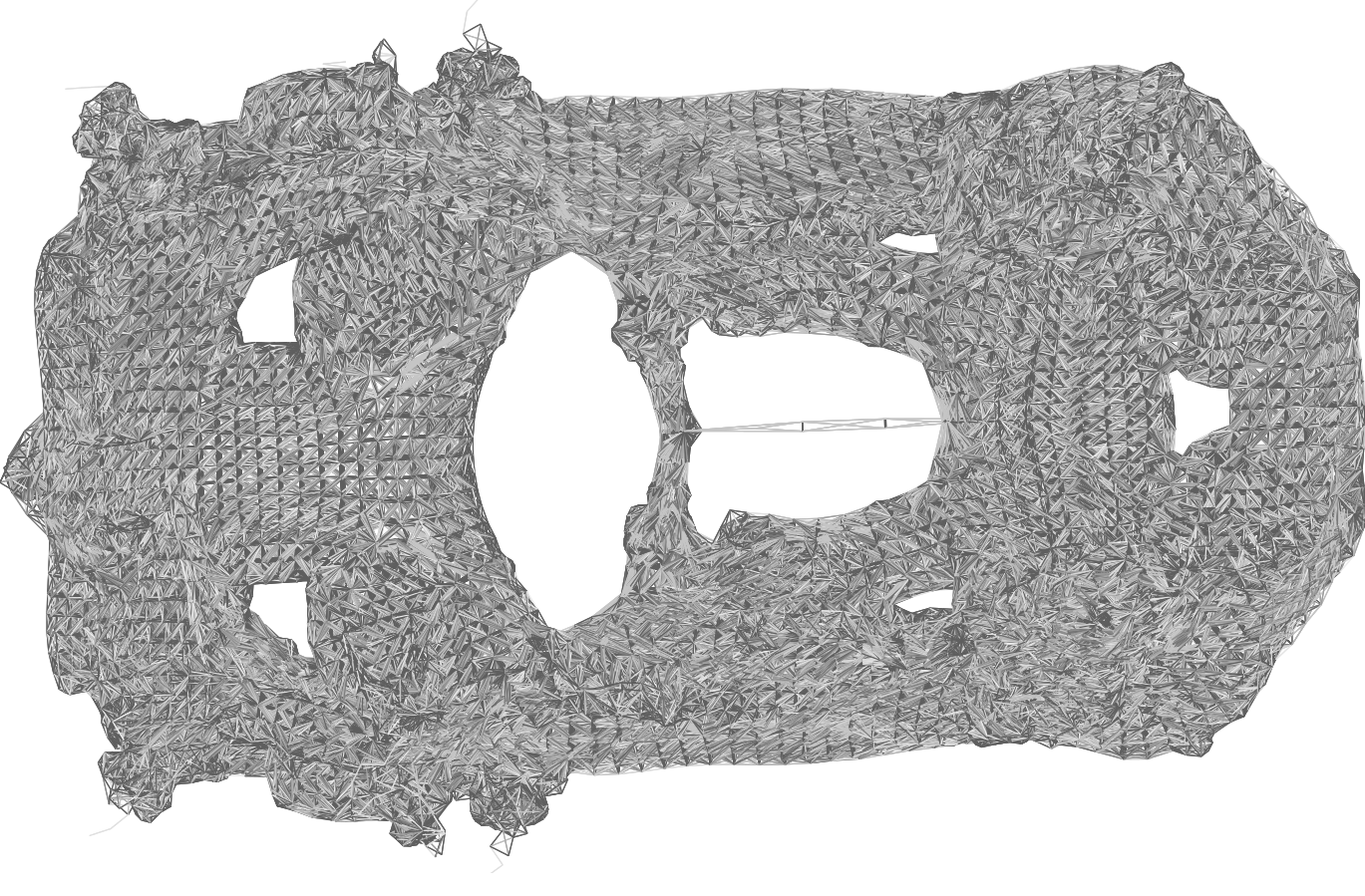} 
\includegraphics[width=3.45cm]{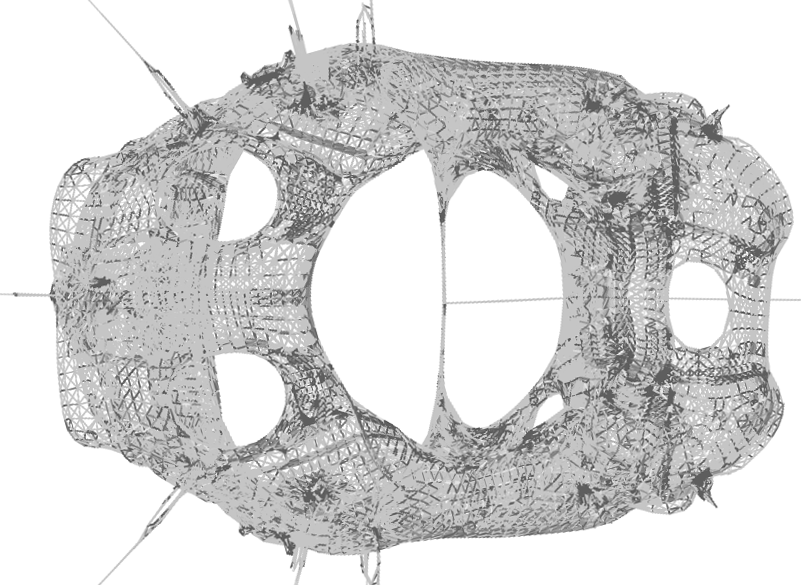}

\caption{Drawings of bcsstk31. Left to right: PivotMDS~\cite{bp-empmsld-07}, Maxent~\cite{ghn-mmgl-13}, MulMent (new).}
\vspace*{-.5cm}
\label{fig:drawingsbyalgorithms}
\end{figure}

Our experimental results in Section~\ref{sec:exp} first reveal that force approximation rarely
affects the layout \emph{quality} significantly -- in terms of maxent-stress values as well as visual quality, 
also see Figure~\ref{fig:drawingsbyalgorithms} and Appendix. The parallel implementation of our multilevel algorithm
 \algoname\  with force approximation is, however, on average 30 times faster than the reference 
implementation~\cite{ghn-mmgl-13} -- and even our sequential approximate algorithm is faster than the reference.
A contribution besides higher speed is that, in contrast to~\cite{ghn-mmgl-13}, our approach does not require
input coordinates to optimize the maxent-stress measure.

\section{Preliminaries}
\label{sec:prelim}

\subsection{Basic Concepts}
Consider an undirected, connected graph $G=(V,E,c,\omega,d)$
with node weights
$c: V \to \MdR_{\geq 0}$, edge weights $\omega: E \to \MdR_{\ge 0}$, target edge lengths $d: E \to \MdR_{>0}$, $n = |V|$, and $m = |E|$.
Often the function $d$ models the required distance between two adjacent vertices.
By default, our initial inputs will have unit edge length $d\equiv 1$ as well as unit node weights and edge weights $c \equiv 1$, $\omega\equiv 1$. 
However, we will encounter weighted problems in the course of our multilevel algorithm.
Let $N(v)\Is \setGilt{u}{\set{v,u}\in E}$ denote the set of neighbors of $v$.
A clustering of a graph is a set of \emph{blocks} (= clusters) of nodes $\{V_1,\dots,V_k\}$ 
that partition $V$, i.e., $V_1\cup \dots \cup V_k=V$ and $V_i\cap V_j=\emptyset$
for $i\neq j$.
A \emph{layout} of a graph is represented as a coordinate vector $x$, where $x_v$ is the two-dimensional coordinate of vertex $v$. 
Since edges are drawn as straight-line segments between their incident nodes, $x$ is sufficient to define the complete graph layout.

\subsection{Related Work}
\label{sec:related}
Most general-purpose layout algorithms for arbitrary undirected graphs are based on physical analogies and can be grouped, according to Hu and Shi~\cite{hs-vlg-15}, into two main classes: algorithms in the \emph{spring-electrical model} and algorithms in the \emph{stress model}.
Both classes of algorithms often yield aesthetically pleasing graph layouts that emphasize symmetries and avoid edge crossings at least in sparse graphs.
Recent surveys of algorithms in these models are given by Hu and Shi~\cite{hs-vlg-15} and by Kobourov~\cite{k-fda-13}.

In the spring-electrical model, first presented by Eades in 1984~\cite{e-hgd-84}, the analogy is to represent nodes as electrically charged particles that repel each other while edges are represented as springs exerting attraction forces to adjacent nodes. 
A graph layout is then seen as a physical system of forces and the goal is to find an optimal layout corresponding to a minimum energy state. 
Spring-electrical algorithms are also known as \emph{spring embedders}, with the algorithm by Fruchterman and Reingold~\cite{fr-gdfp-91} being one of the most widely used spring embedder algorithms. 
It simulates the physical system of attractive and repulsive forces and iteratively moves each node into the direction of the resulting force.
Each iteration requires, however, a quadratic number of force computations due to the repulsive forces between all pairs of nodes, which limits the scalability of the original approach.
A faster approximative force calculation method based on quadtrees, aggregating especially the long-range forces, 
has been proposed by Barnes and Hut~\cite{barnes1986hierarchical} and yields running times of $O(n \log n)$ under certain assumptions.

The (full) stress model is closely related to multidimensional scaling~\cite{k-msognh-64}, and was introduced in graph drawing by Kamada and Kawai~\cite{kk-adgug-89}.
It is based on defining ideal distances $d_{uv}$ not only between adjacent vertices but between all vertex pairs $(u,v) \in V \times V$ and then minimizing the layout stress $\sum_{u \ne v} w_{uv} (||x_u - x_v|| - d_{uv})^2$, where $w_{uv}$ is a weight factor typically chosen as $w_{uv} = 1/d_{uv}^2$.
Often, the distance $d_{uv}$ between adjacent nodes is set to 1, while the distance of non-adjacent nodes is the shortest-path distance in the graph.
Solving this model is typically done by iteratively solving a series of linear systems~\cite{gkn-gdsm-05}.
The need to compute all-pairs shortest paths and to store a quadratic number of distances again defeats the scalability of this original approach for large graphs.
One of the fastest algorithms for approximatively solving the stress model instead is PivotMDS~\cite{bp-empmsld-07}, which requires distance calculations from each vertex only to a small set of $k \ll n$ suitably chosen pivot vertices.
 
The stress model prescribes target distances not only for edges but for all vertex pairs.
While this is a reasonable approach, it still brings artificial information into the layout process.
An interesting alternative has been proposed by Gansner et al.~\cite{ghn-mmgl-13}. 
Their algorithm (called \emph{Maxent}) uses the sparse stress model, which only contains the stress terms for the edges of the graph.
In order to deal with the remaining degrees of freedom in the layout, they suggest using the maximum entropy principle instead.
Since our algorithm is closely related to Maxent, we discuss the latter in more detail in Section~\ref{par:maxent}.
 
A general approach for speeding up layout computations for large graphs is the \emph{multilevel technique}, which has been used in the spring-electrical~\cite{w-mafg-03,qe-fgdcva-01} and in the stress model~\cite{gk-ggdwip-02}. 
A multilevel algorithm computes a sequence of increasingly coarse but structurally related graphs as abstractions of the original graph.
Starting from a layout of the coarsest graph, incremental refinement steps using the previous layout as a scaffold eventually produce a layout of the entire input graph, where the refinement steps are fast due to the good initial layouts. 

In addition to sequential algorithms for drawing large graphs, there is previous research in parallel layout algorithms, particularly using a graphics processing unit (GPU). 
Frishman and Tal~\cite{ft-mgl-07} presented a multilevel force-based layout algorithm and implemented it using GPU-based parallelization.
Ingram et al.~\cite{imo-gm-09} also exploit parallel GPU computations and presented a multilevel stress-based layout algorithm.

\subsection{Maxent-stress optimization}
\label{par:maxent}

Gansner et al.~\cite{ghn-mmgl-13} proposed the maxent-stress  model that combines a sparse stress model with an entropy term to resolve the degrees of freedom for non-adjacent vertex pairs. 
The entropy term itself is optimized when all nodes are spread out uniformly, similar to the repulsive forces in the spring-electrical model.
Gansner et al.~\cite{ghn-mmgl-13} showed that the maxent-stress model performs well on several measures of layout quality in distance-based embeddings and avoids typical shortcomings of other stress models, particularly for non-rigid graphs. 
Formally, the maxent-stress $M(x)$ of a layout $x$  is defined\footnote{In fact, Gansner et al.\ define a slightly more general model that considers the stress term for arbitrary supersets $S \supseteq E$ and allows variations of the entropy term. Our algorithm also works for the general model; to simplify the description, we restrict ourselves to the default model.}
as
\begin{equation}
\label{eq:maxent-stress}
	M(x) = \sum_{\{u,v\}\in E} w_{uv} (||x_u - x_v|| - d_{uv})^2 - \alpha \sum_{\{u,v\}\not\in E} \ln ||x_u -x_v||,
\end{equation}
where $d_{uv}$ is the target distance between nodes $u$ and $v$ and $w_{uv}$ is a weight factor typically chosen as $w_{uv} = 1/d_{uv}^2$.
Throughout the paper, we use this as a weight factor.
The scaling factor $\alpha$ is used to modulate the strength of the entropy term and is gradually reduced in the implementation.

Gansner et al.\ minimize the maxent-stress using a technique  
that repeatedly solves Laplacian linear systems that additionally include a repulsive force vector which is approximated following the quadtree method of Barnes and Hut~\cite{barnes1986hierarchical}.

Alternatively, they proposed (but did not implement) the following local iterative force-based scheme to solve the maxent-stress model:
\begin{equation}
	x_{u} \gets \frac{1}{\rho_{u}} \sum_{\{u,v\}\in E} w_{uv} \left(x_{v} + d_{uv} \frac{x_{u}-x_{v}}{\Vert x_{u}-x_{v}\Vert}\right) + \frac{\alpha}{\rho_{u}} \sum_{\{u,v\}\notin E} \frac{x_{u}-x_{v}}{\Vert x_{u}-x_{v}\Vert^{2}},
\label{eq:stress-entropy-iter}
\end{equation}
where $\rho_{u}=\sum_{\{u,v\}\in E}w_{uv}$. Note that sometimes we use the abbreviation $r(u,v):= \frac{x_{u}-x_{v}}{\Vert x_{u}-x_{v}\Vert^{2}}$ and shortly call these values \emph{$r$-values}.

\section{Multilevel Maxent-stress Optimization}
\label{sec:algo}
As mentioned, a successful (meta)heuristic for graph drawing (and other optimization problems on large graphs) is the multilevel approach. We employ it for maxent-stress optimization also for other reasons:
(i) Some graphs (such as road networks) feature a hierarchical structure, which can be exploited to some extent
by a multilevel approach and (ii) the computed hierarchy may be useful later on for multiscale visualization.

Before going into the details, we briefly sketch our algorithmic 
approach: The method for creating the graph hierarchy is based on fast graph clustering with 
controllable cluster sizes. 
Each cluster computed on one hierarchy level is contracted into a new supervertex for the next level.
After computing an initial layout on the coarsest hierarchy level, we improve the drawing 
on each finer level by iterating Eq.~(\ref{eq:stress-entropy-iter}). Additionally, this process exploits
the hierarchy and draws vertices that are densely connected with each other (\ie which are in the same cluster) close to each other.

\subsection{Coarsening and Initial Layout}
\label{sub:coarsening}
To compute the clustering that is contracted in the manner described above, we adapt size-constrained label propagation
(SCLaP)~\cite{pcomplexnetworksviacluster}, an algorithm originally developed for coarsening and local improvement during
multilevel graph partitioning. SCLaP itself is based on the graph clustering algorithm \emph{label 
propagation}~\cite{labelpropagationclustering}. The latter starts with a singleton clustering (\ie each node is a cluster).
The algorithm then works in rounds. Roughly speaking, in each round the algorithm visits all nodes in random order and assigns 
each node to the predominant cluster in its neighborhood (illustrated in Figure~\ref{fig:lp} in the appendix). 
This way, cluster IDs (= labels) propagate through the graph and nodes in a dense cluster usually agree on a common label. 

However, clusters with unconstrained sizes are not desirable here since they would hamper convergence of the local improvement 
phase. The trade-off between this convergence speed and the number of hierarchy levels needs to be chosen
properly for a fast overall running time. That is why SCLaP constrains cluster sizes,
\ie it introduces an upper bound $U := \max( \max_v c(v), W)$ on the cluster sizes ($W$ is specified below).
Consequently, in each SCLaP round, nodes are assigned to the predominant cluster that is not overloaded 
after the label change.

In our implementation, based on preliminary experiments, we set the parameter $W$ to $\min(b^h,\frac{|V|}{f})$, where $b$ and $f$ are tuning parameters and $h$ is the level in the hierarchy that we are currently working on. The intuition behind this choice is that we want the contraction process not to be too strong on the fine levels in order two allow fast convergence of local improvement algorithms, whereas we allow stronger contractions on coarser levels. 
If the contracted graph is not more than 10\% smaller than the graph on the current level, we decrease the value of $f$ 
and set it to~$0.7f$. 

While the original label propagation algorithm repeats the process until convergence, SCLaP performs at most $\ell$ rounds, 
where $\ell$ is a tuning parameter. One round  of the algorithm can be implemented to run in 
$\Oh{n+m}$ time.

Contracting a clustering works as follows: 
each block of the clustering is contracted into a single node. 
The weight of the node is set to the sum of the weight of all nodes in the original block. 
There is an edge between two nodes $u'$ and $v'$ in the contracted graph if the
two corresponding blocks in the clustering are adjacent to each other in $G$,
\ie block $u'$ and block $v'$ are connected by at least one edge.
The weight of an edge $(u',v')$ is set to the sum of the weight of edges that run between block $u'$ and block $v'$ of the clustering. 
An example contraction is shown in Figure~\ref{fig:clustercontraction} in the appendix. 

\paragraph*{Initial Layout.}The process of computing a size-constrained clustering and contracting it is repeated recursively. 
Then an initial layout is drawn, meaning that each of the two nodes of the coarsest graph is assigned to a position. 
We place the vertices such that the distance is optimal. The optimal distance of the two vertices is defined and motivated in the next section. 

\subsection{Uncoarsening and Local Improvement}
\label{sub:refinement}
When the initial layout has been computed, the solution is successively prolongated to the next finer
level, where a local maxent-stress minimizer is used to improve the layout.
For undoing the contraction, nodes that have been in a cluster are drawn at a random position around the location of its coarse representative. More precisely, let $v$ be a (fine) vertex that is represented by the coarse supervertex $v'$ at $P=(x,y)$. We place $v$ at a random position in a circle around $P$ with radius $r:=\sqrt{c(v')}$. We do this by picking an angle uniformly at random in $[0,2\pi]$ and a distance to $P$ uniformly at random in $[0,r]$. These two values are then used as a polar coordinate for $v$ with respect to the origin $P$.

\paragraph*{Local Improvement.}
Our local improvement tries to minimize the maxent-stress on each level of the hierarchy
based on Eq.~(\ref{eq:stress-entropy-iter}). Note, however, that simply iterating Eq.~(\ref{eq:stress-entropy-iter})
on each level is not sensible since coarse vertices represent a multitude of vertices. These vertices need space to be drawn on the next finer level. Now let $u$ and $v$ be two vertices on the same fixed level.
We adjust distances $d_{uv}$ on the current level in the hierarchy under consideration to $\sqrt{c(u)}+\sqrt{c(v)}$ with the intuition that vertices represented by $u$ should be drawn in a circle around $u$ with radius $\sqrt{c(u)}$ (similarly for~$v$). 

As Gansner \etal~\cite{ghn-mmgl-13}, we adjust the value of $\alpha$ in Eq.~(\ref{eq:stress-entropy-iter})
during the process. Since we want to approximate the maxent-stress, the value should be small. However, it cannot be too small initially since one would only solve a sparse stress model in this case.
Hence, following Gansner \etal~\cite{ghn-mmgl-13}, we set $\alpha$ to one initially and gradually reduce it by $\alpha := 0.3 \cdot \alpha$ until $\alpha_{\min}=0.008$ is reached.

We call a single update step of the coordinates of all vertices using Eq.~(\ref{eq:stress-entropy-iter}) an
\emph{iteration}. Multiple iterations with the same value of $\alpha$ are called \emph{round}. 
The current iteration uses the coordinates that have been computed in the previous iteration.
We perform at most $a$ iterations with the same value of $\alpha$ in one round. Then we reduce $\alpha$ as described above.
If the relative change $||x^{\ell+1}-x^\ell||/||x^\ell||$  in the layout  is smaller than some threshold $\epsilon$, we directly reduce the value of $\alpha$ and continue with the next round.

\paragraph*{Faster Local Improvement.}
The local optimization algorithm presented above has a theoretical running time of $\mathcal{O}(n^2)$ per iteration.
To speed this up, one can use approximations for the distances in the entropy term in Eq.~(\ref{eq:stress-entropy-iter}).
We do this by taking the cluster structure computed during coarsening into account:
Let  $V_1\cup \ldots \cup V_k$ be the corresponding clustering and $M: V \to V'=\{1, \ldots, k\}$ be the mapping that maps a node $v \in V$ to its coarse representative.
The first term in Eq.~(\ref{eq:stress-entropy-iter}) is computed as before and the second term is approximated by using the coordinates of the corresponding coarse vertex.
As formula the second term written without the multiplicative factor $\frac{\alpha}{\rho_{u}}$ becomes 
\begin{equation}
\sum_{u\neq v \atop M(u) = M(v)}r(u,v)+
\sum_{v' \in V' \atop v' \neq M(u)}\nu(v')\frac{x_{u}-x'_{v'}}{\Vert x_{u}-x'_{v'}\Vert^{2}}-
\sum_{\{u,v\}\in E} r(u,v),
\label{eq:fasterstress}
\end{equation}
where $x'$ maps a coarse vertex to its coordinates and $\nu(v')$ is the \emph{number of nodes} that the coarse vertex represents on the \emph{current} finer level. 
Roughly speaking, we reduced the necessary amount of computation to add up the values of $r$ by summing up the correct values of $r$ for all vertices that are in a sense \emph{close} and using approximations for vertices that are far away. In our context, a vertex is close if it is in the same cluster as the currently processed vertex. 
If a vertex is not close, we use the coordinate of its coarse representative instead. 
We avoid unnecessary computation by scaling the approximated value of $r$ with the number $\nu(v')$ of vertices it represents and adding approximated value of $r$ only once.
The last term in Eq.~(\ref{eq:fasterstress}) subtracts values of $r$
for $\{u,v\} \in E$ that have been added in good faith in the first two summations.

Note that if $M$ is the identity, then the term in Eq.~(\ref{eq:fasterstress}) is the same as in the original Eq.~\ref{eq:stress-entropy-iter}. In this case the first two summations add up the $r$-values  for all pairs of vertices and the last sum subtracts the $r$-values for pairs that are in $E$. 

After the update of the vertices on the current level, we update the coordinates of the vertices on the coarser level used for approximation.
We set the coordinate of a vertex $v'$ on the coarser level to the weighted midpoint of the vertices represented by~$v'$. 

Note that one obtains even faster algorithms by using a coarser version of the graph that is \emph{multiple levels beneath} the current level in the graph hierarchy. 
That means instead of using the next coarser graph, we use the contracted graph which is $h>1$ levels beneath the current graph in the hierarchy -- if there is such. 
Otherwise, we use the coarsest graph in the hierarchy. 
Obviously this yields a trade-off between solution quality and running time.
Also note that this introduces an additional error. To see this, let the coarser vertices that have the same coarse representative on the level used for approximating values of $r$ be called $\mathcal{M}$-vertices (merged vertices). Now, for a vertex on the current level, the $r$-values of $\mathcal{M}$-vertices are not accounted for in Eq.~(\ref{eq:fasterstress}). 
Hence, we look at the parameter $h$ carefully in Section~\ref{s:experiments} and evaluate its impact on running time and solution quality.
We call our algorithms \algorithmname\  and  denote by $\text{\algorithmname}_h$ the  algorithm that uses an $h$-level approximation of the $r$-values. With $h$ if $h=0$ we denote the quadratic-time
algorithm. A rough analysis in Appendix~\ref{sec:running-time} yields:

\begin{proposition}
Under the assumption of equal cluster sizes, the running time of one iteration of algorithm
 \algoname$_{h}$, $h \geq 0$, is $\mathcal{O}(m+  n^{\frac{h+2}{h+1}})$, respectively.
\end{proposition}

Properly implemented, multilevel algorithms lead to fast convergence of their local optimizers. 
Moreover, the overall work performed by the multilevel approach is only a constant factor times the one on the finest level.
This leads us to the initial appraisal that the same asymptotic running times may hold for the respective complete algorithms.

\paragraph{Shared-memory Parallelization.}
Our shared-memory parallelization of an iteration of the local optimizer uses OpenMP and works as follows: Since new coordinates of the vertices in the same iteration can be computed independently, we use multiple threads to do so. The relative change in the layout $||x^{\ell+1}-x^\ell||/||x^\ell||$ can be computed in parallel using a reduce operation.
Parallelism is also used analogously when working on different levels for the distance approximations
in the entropy term.
Other parts of the overall algorithm could potentially be parallelized, too -- such as coarsening. However, already on medium sized graphs coarsening consumes less than 5\% of the algorithm's overall running time. Moreover, the relative running time of coarsening decreases even more with increasing graph size so that the effort does not seem worth it.

\section {Experimental Evaluation}
\label{sec:exp}
\label{s:experiments}

\paragraph{Methodology.}
We implemented\footnote{We will release the implementation of our algorithms as open source.} the algorithm described above using C++.  
Parallelization of our algorithm has been done using OpenMP.
We compiled our programs using g++ 4.9 -O3 and OpenMP 3.1.  
Executables for PivotMDS (PMDS)~\cite{bp-empmsld-07} and MaxEnt (GHN, for clarity we
use the author names as acronym)~\cite{ghn-mmgl-13} have been kindly provided by Yifan Hu. 
When comparing layouts computed by different algorithms, we evaluate two metrics. The first metric is the full stress measure, $F(x) = \sum_{u,v\in V}w_{uv}(||x_u-x_v||-d_{uv})^2$, and the second one is the maxent-stress function $M(x)$ as defined in Eq.~(\ref{eq:maxent-stress}) at the final penalty level of $\alpha=0.008$.
The latter is of primary importance since that is what GHN and \algoname\ optimize for.
The implementations PMDS and GHN sometimes compute vertices that are on the same position. Hence, we add small random noise to the coordinates of these layouts in order to be able to compute the maxent-stress.
More precisely, for each of the components of the 2D-coordinate of a node, we randomly add or subtract a random value from the interval $[10^{-7},10^{-4}]$. This changes the full stress measure by less than $10^{-4}$ percent on average.
We follow the methodology of Gansner~\etal~\cite{ghn-mmgl-13} and scale the layout of all algorithms to minimize the stress to be fair to all methods:
We find a scalar $s$ such that $\sum_{u,v \in V}w_{uv}(s||x_u - x_v|| - d_{uv})^2$ is minimized for a given layout~$x$. 

\paragraph{Machine.} Our machine has four Octa-Core Intel Xeon E5-4640 (Sandy Bridge) processors (32 cores, 64 with hyperthreading active) which run at a clock speed of 2.4 GHz. 
It has 512 GB local memory, 20 MB L3-Cache and 8x256 KB L2-Cache. 
Unless otherwise mentioned, our algorithms use all 64 cores (hyperthreading) of that machine.   Since PMDS and GHN are sequential algorithms, they use one core of that machine.

\paragraph{Algorithm Configuration.}
After an extensive evaluation of the parameters, we fixed the cluster coarsening parameters $f$ to 20 and $b$ to 2. The initial value of the penalty parameter $\alpha$ is set to 1. We perform at most $a=2$ iterations with the same value of $\alpha$, while it has not reached its minimum value of 0.008. 
When it has reached its minimum value, we iterate until the relative error $||x^{\ell+1}-x^\ell||/||x^\ell||$ is smaller than $0.0001$. 
Yet, our experiments indicate that our algorithm is not very sensitive about the choice of these parameters. 
We evaluate the influence of the approximation level $h$ in Section~\ref{s:influenceofh}. 
\paragraph{Instances.}
\label{sub:settings}
We use the instances \Id{1138\_bus}, \Id{USpowerGrid}, \Id{bcsstk31}, \Id{commanche} and \Id{luxembourg} employed in \cite{ghn-mmgl-13} and extend the set to include larger instances.
We excluded the graphs \Id{gd}, \Id{qh882} and \Id{lp\_ship04l} from \cite{ghn-mmgl-13} from our experiments since the graphs are either not undirected or the corresponding matrix is rectangular.
Most of the instances taken from~\cite{ghn-mmgl-13} are available at the Florida Sparse Matrix Collection~\cite{UFsparsematrixcollection}.
The graphs \Id{3elt}, \Id{bcsstk31}, \Id{fe\_pwt} and \Id{auto} are available at the Walshaw benchmark archive~\cite{soper2004combined}.
The graphs \Id{del$X$} are Delaunay triangulations of $2^{X}$ random points in the unit square~\cite{kappa}.  
Moreover, the graphs \Id{nyc} and \Id{luxembourg} are road networks. 
These graphs have been taken from the benchmark set of the 9th and 10th DIMACS Implementation Challenge~\cite{demetrescu2009shortest,benchmarksfornetworksanalysis}.
A summary of the basic properties of these instances can be found in Table~\ref{tab:graphstable} in the appendix.
In any case, we draw the largest connected component if the graph has more than one.
We assume unit length distance for all graphs.

\subsection{Influence of Coarse Graph Approximation and Scalability}
\label{s:influenceofh}
\label{s:scalability}
\label{sp:scalability}
\begin{figure}[t]
\vspace*{-.75cm}
\centering
\includegraphics[width=6cm,page=1]{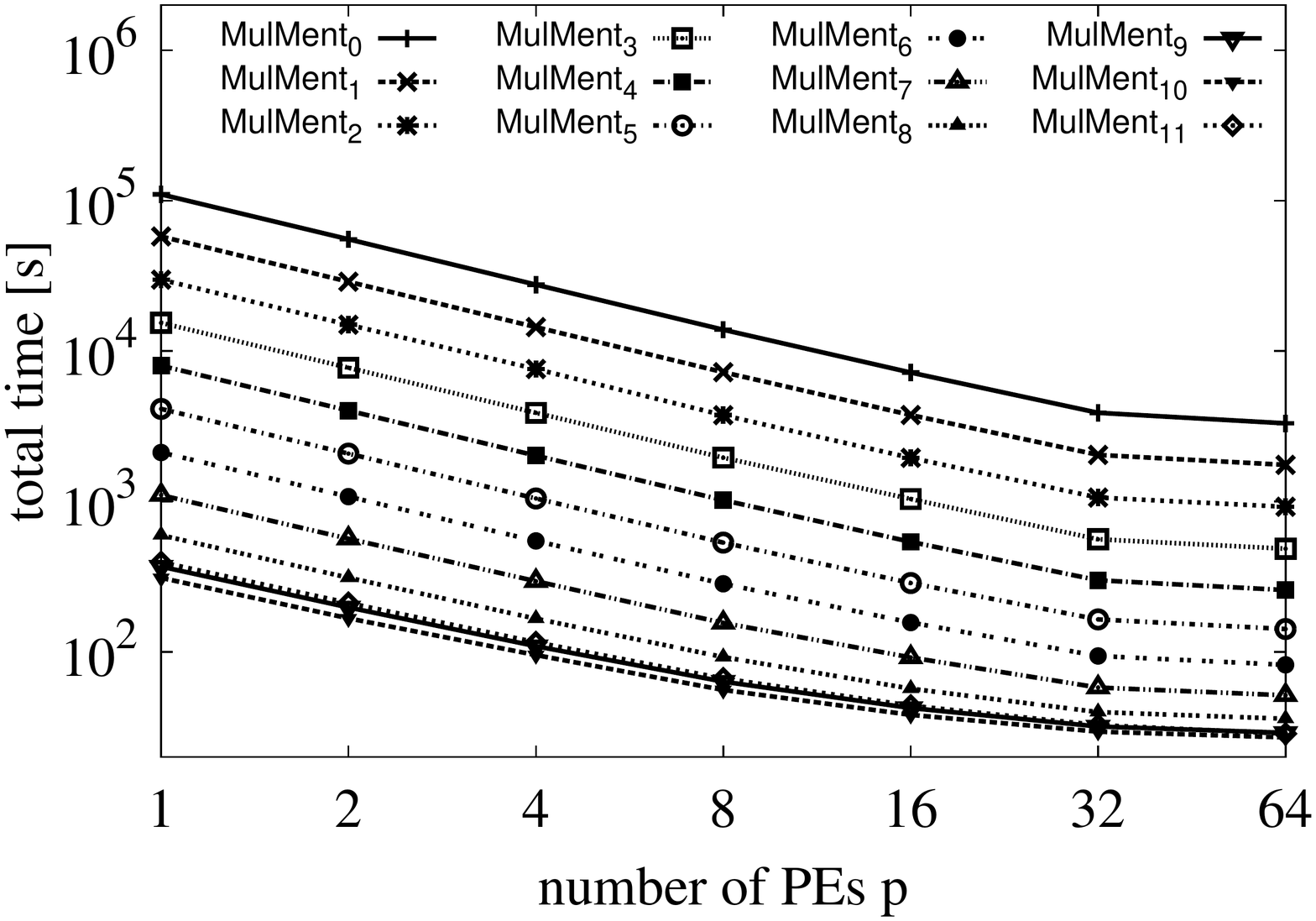}  \includegraphics[width=6cm,page=2]{plots/scalability.pdf} \\
\vspace*{-.5cm}
\caption{Running times and parallel speedups of our algorithms on $\Id{del20}$.}
\vspace*{-.5cm}
\label{fig:speedupplots}
\end{figure}

In this section, we investigate the influence of the parameter $h$ on layout quality and running time (algorithmic speedup) as well as the scalability of our algorithms with varying number of threads (parallel speedup).
We perform detailed experiments on our medium sized networks (using 64 threads) and present parallel speedups on the largest graphs \Id{auto} and \Id{del20}. 
We report absolute running times and parallel speedups for the graphs \Id{auto} and \Id{del20} in Figures~\ref{fig:speedupplots} and \ref{fig:speedupplotsauto} and present detailed data for the medium size networks in  Table~\ref{tab:influenceh} in the appendix. 
We do not report layout quality metrics for \Id{auto} and \Id{del20} since the size of the network makes it infeasible to compute them and 
the result of the algorithm is independent of the number of threads used.

We now investigate the influence of the parameter $h$. In general, the larger the graphs get, the larger the algorithmic speedups obtained with increasing $h$.
On the smallest graph in this collection, we obtain an algorithmic speedup of about 3 with $h=6$ (\Id{fe\_pwt}) over \algorithmname$_0$. On the largest two instances in this section, we obtain an algorithmic speedup of 30 with $h=9$ (\Id{auto}) and of 122 with $h=10$ (\Id{del20}). 
In addition, the precise choice of the parameter does not seem to have a very large impact on solution quality on these graphs. This is also due to the size of the networks.
As apparent from Tables~\ref{tab:influenceh} and~\ref{tab:compmaxentstress} in the appendix, the graphs on which full stress measure slightly increases are luxembourg and bcsstk31 (7\% and 15\% respectively). The metric actually under consideration, maxent-stress, always remains comparable.
On all instances under consideration, we observe a locally optimal value for $h$ in terms of running time. It is around seven and seems to get larger with increasing graph size. This is due to the fact that too large values of $h$ provide less precision and slower~convergence.

On \Id{del20}, the scalability with the number of threads is almost perfect for small values of $h$. With enabled hyperthreading, we achieve slightly \emph{superlinear} speedups for MulMent$_0$. 
As less work has to be done for increasing $h$, speedups get smaller. 
The smallest speedup on this graph has been observed for MulMent$_{10}$. 
In this case, we achieve a speedup of 11.5 using 64 threads over MulMent$_{10}$ using one thread. 
With even larger $h$ speedups increase again. 
The parallel scalability on \Id{auto} is similar.

Another interesting way to look at the data is the overall speedup -- algorithmic and parallel speedup combined -- achieved over MulMent$_0$ using only one thread. 
The largest overall speedup is obtained by \algorithmname$_{10}$ using 64 threads. In this case, the overall speedup  is larger than 4000~--~reducing the running time of the algorithm from 30~hours to 27~seconds. Speedups over PMDS and GHN are found in the next section.

\subsection{Comparison to other Drawing Algorithms}
\label{s:comparsion}
\label{sub:results}
We now compare \algoname\ to the two implementations PMDS~\cite{bp-empmsld-07} and GHN~\cite{ghn-mmgl-13}. 
We do this on all networks but only report quality metrics for small and medium sized graphs since it is infeasible to compute quality metrics for the large graphs. We report detailed data in Tables~\ref{tab:compmaxentstress} and~\ref{tab:comprunningtime} in the appendix.

Most importantly, although \algoname\ sometimes performs a few
percent worse than GHN, the maxent-stress of all layouts is more or less \emph{similar}. 
PMDS performs slightly worse in this metric.
Intriguingly, the alternative full stress metric is consistently better on small networks 
for \algoname\ than the results obtained by PMDS (except for $h=10$).
On the other hand, full stress obtained by our algorithms is comparable to the layout computed by GHN on four out of nine instances.  
On the three largest medium sized networks, we obtain worse full stress than PMDS and GHN. 
However, this is not astonishing since our algorithm does not optimize for full stress -- 
in \emph{contrast} to PMDS. And GHN at least starts with a PMDS solution
and improves maxent-stress afterwards.

Our implementations of \algorithmname$_{7,10}$ are always faster than GHN, both of them a factor 30 on average.
Also, \algorithmname$_{7,10}$ outperform even PMDS in terms of running time as soon as the graphs get large enough (medium and large sized graphs).
On the large graphs, \algorithmname$_{10}$ is a factor of 2 to 3 faster than PMDS and a factor of 32 to 63 faster than GHN. 
In addition, \algorithmname$_{7,10}$ are also several times faster than GHN when using one thread only (see Appendix~Table~\ref{tab:comprunningtimesinglethread}).

\subsection{Dynamic Networks}
\label{s:dynamicnetworks}
One of the main advantages of the iterative scheme is its ability to use an existing layout 
for computing a new one, \eg for a graph that has changed over time.
We perform experiments with dynamic graphs obtained by modifying our medium sized networks.
Often one is interested in drawing graphs that have more or less good locality.
Hence, we define a random model that modifies the edges of a graph by removing random edges and inserting edges between vertices that are not too far apart. 

To be more precise, we start with an input graph $G$ and perform a breadth first search from a random start node to compute a random spanning tree. 
We then remove $x$\% undirected non-tree edges at random in the beginning. Note that this ensures that the graph stays connected.
Afterwards, we insert $x$\% new edges as follows. We pick a random node and insert an undirected edge to a random node that has distance $1<d\leq\mathcal{D}$ in the original graph $G$, where $\mathcal{D}$ is a tuning parameter. We denote the graph that results out of this process as $Q$.

We compute two layouts of $Q$. The first one updates coordinates given by an initial layout of $G$ (update algorithm). The second layout is computed by our algorithm from scratch (scratch algorithm), \ie discarding the initial layout.
In the first case, we start directly at the penalty level $\alpha=0.008$ and only update coordinates on the finest level of the hierarchy. 
We compute the graph hierarchy as before but stop the coarsening process after the computation of $h$ levels. 
Coordinates of the vertices on the approximation level are set to the middle point of the vertices in the corresponding cluster initially. 

We vary $x\in \{1,5\}$, $\mathcal{D} \in \{2,16\}$ and $h \in \{0,7\}$, and present detailed data in Table~\ref{tab:dynamicresults} in the appendix.
As expected, the \emph{running time} of the update algorithm ($t_\text{dyn}$) is always smaller than the running time of the scratch algorithm ($t_{\text{scratch}}$). 
As MulMent$_7$ performs less work than MulMent$_0$, algorithmic speedups are always larger for the latter. 
For $h=0$, the update algorithm is a factor of 4 faster than the scratch algorithm on average.
On the other hand, for $h=7$ the update algorithm saves about 50\% time on average over the scratch algorithm. 
\emph{Solution quality} is not influenced much. On average, the full stress measure of the update algorithm is 9\% larger and maxent-stress improves by~1\% compared to the scratch algorithm. The increase in full stress is mostly due to the Delaunay instance and $\mathcal{D} = 16$, in which the full stress of the layout of the update algorithm is a factor of two larger. The algorithmic speedup does not seem to be largely influenced by $\mathcal{D}$. However, we expect that much larger values of $\mathcal{D}$ will decrease the speedup of the update algorithm over the scratch algorithm.

\section{Conclusions}
\label{sec:concl}

We have presented a new multilevel algorithm for iteratively and approximatively optimizing the maxent-stress model, a model proposed by
Gansner \etal~\cite{ghn-mmgl-13} to avoid typical pitfalls of other stress models. 
From the experimental evaluation we conclude that our parallel algorithm produces layouts with similar visual quality and maxent-stress values as the reference implementation~\cite{ghn-mmgl-13}. At
the same time it is on average 30 times faster, even more for dynamic graphs. Moreover, our algorithm is even up to twice as fast as the fastest stress-based algorithm PivotMDS~\cite{bp-empmsld-07}. It thus  combines the high speed of PivotMDS with the high visual quality of Maxent in a single algorithm, at least if a multicore system is available.

\vspace*{-.25cm}
{
\subsubsection*{Acknowledgements.}
Financial support by DFG is acknowledged (DFG grants
ME 3619/3-1 and SA 933/10-1). We thank Yifan Hu for providing us the codes from \cite{ghn-mmgl-13}.
}

\vfill
\pagebreak

\begin{appendix}
\section{Additional Figures}
\begin{figure}[htb]
\begin{center}                                                      
                 \includegraphics[width=.75\textwidth]{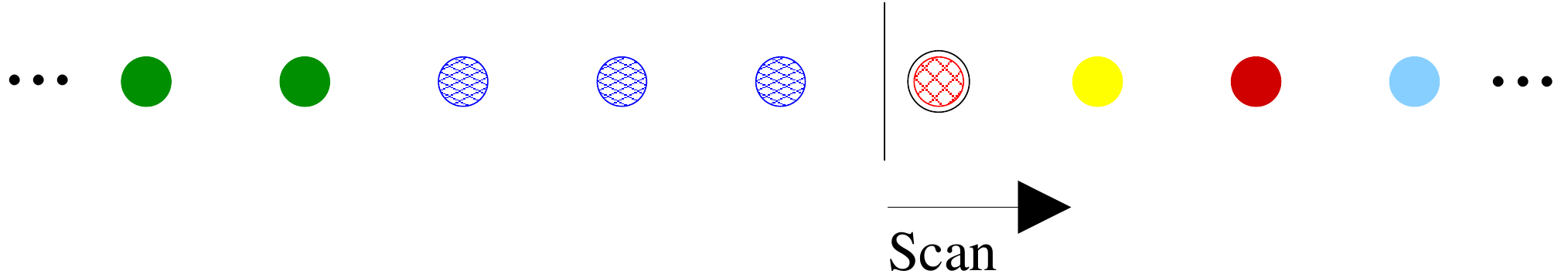}
     \end{center}
     \begin{center}
     \begin{tabular}{ccc}
             \includegraphics[width=3cm]{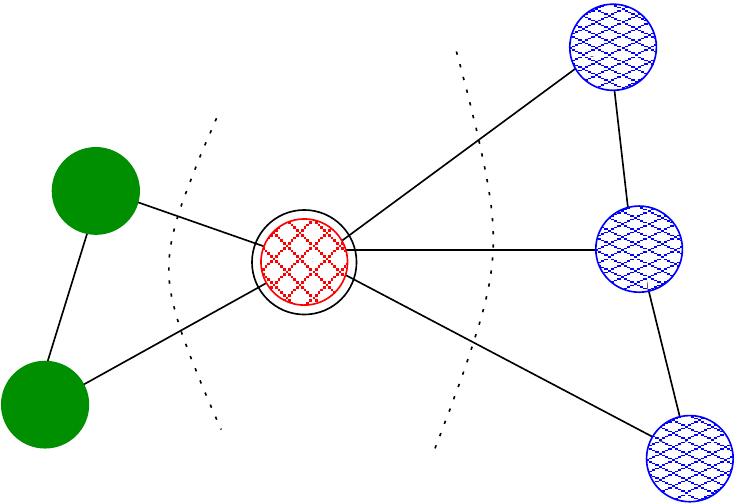} & \begin{minipage}{.025\textwidth}\vspace*{-2cm}\textbf{$\rightarrow$}\end{minipage} &
                \includegraphics[width=3cm]{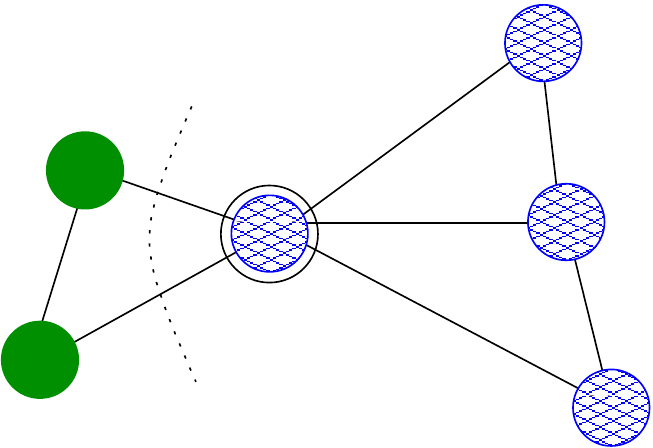}
        \end{tabular}
        \end{center}
        \caption{An example round of the label propagation graph clustering algorithm. Initially each node is in its own block. The algorithm scans all vertices in a random order and moves a node to the block with the strongest connection in its neighborhood.}
        \label{fig:lp}
\end{figure}

\begin{figure}[htb]
\centering
\includegraphics[width=0.5\textwidth]{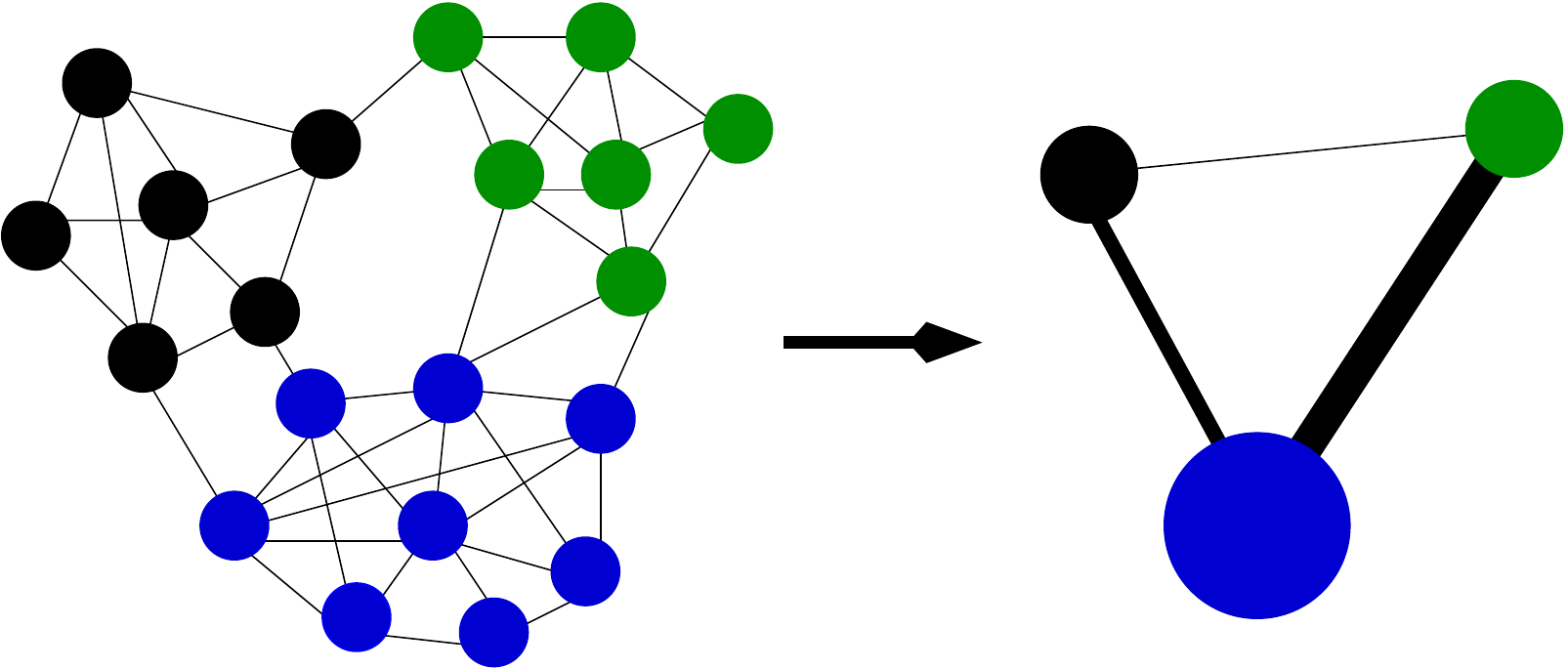}
\caption{Contraction of a clustering (colors indicate cluster labels). Each cluster of the 
graph on the left corresponds to a node (= supervertex) in the graph on the right.}
\label{fig:clustercontraction}
\end{figure}
\begin{figure}[h!]
\vspace*{-.75cm}
\centering
\includegraphics[width=6cm,page=1]{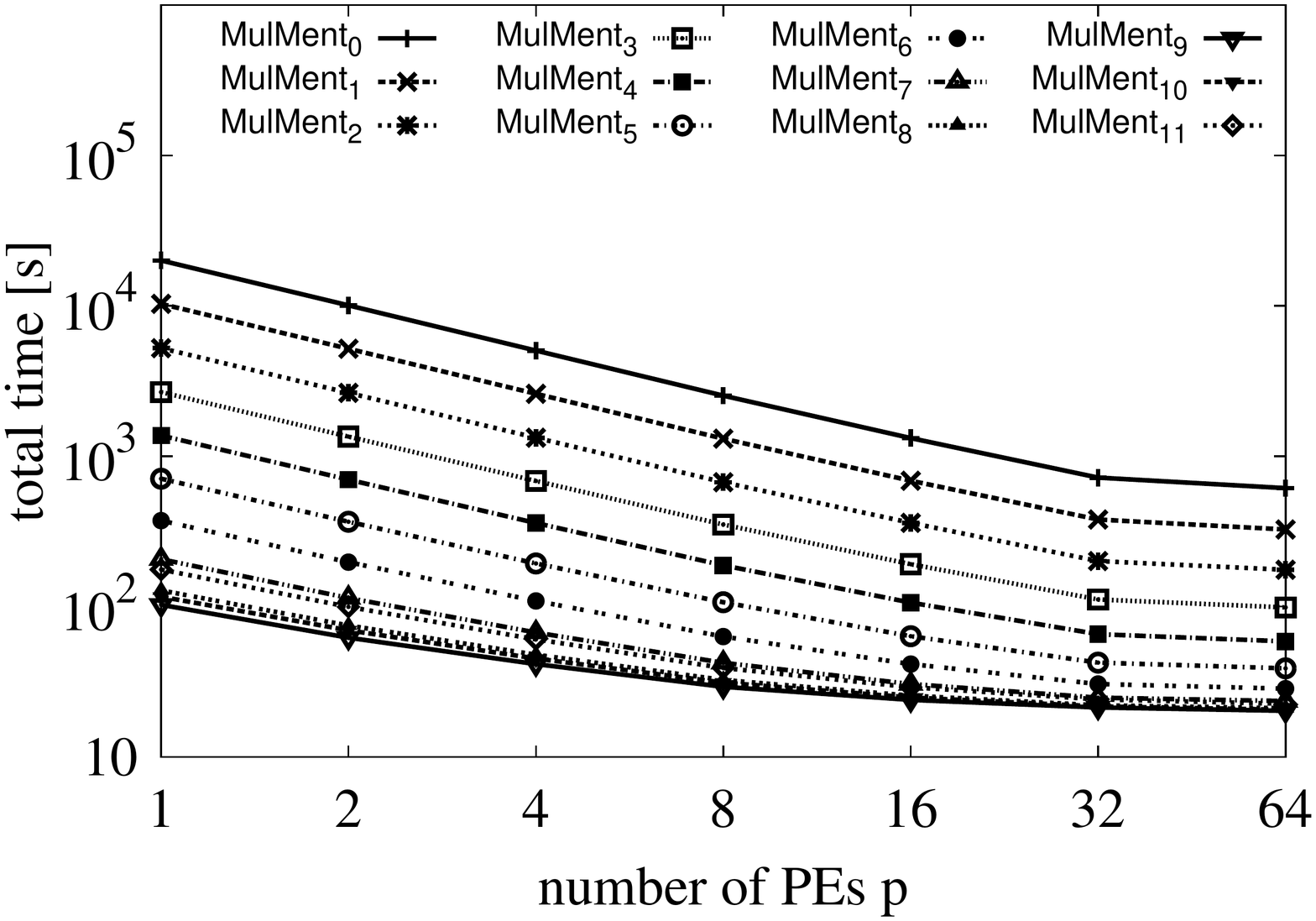}  \includegraphics[width=6cm,page=2]{plots/scalability_auto.pdf} \\
\vspace*{-.5cm}
\caption{Running times and parallel speedups of our algorithms on $\Id{auto}$.}
\vspace*{-.5cm}
\label{fig:speedupplotsauto}
\end{figure}

\vfill
\pagebreak
\section{Running Time Proof Sketch}
\label{sec:running-time}
We start with sketching the running time on the finest level of the hierarchy for $h=1$. 
To simplify the analysis of our algorithm, we assume that our clustering algorithm always
computes clusters of equal size, \ie each cluster contains exactly $c$ vertices. In this
(admittedly not overly realistic case), there are $n/c$ clusters. 
It is easy to see that we need $\mathcal{O}(d(v) + c + n/c )$ time to evaluate Eq.~(\ref{eq:fasterstress}) for a vertex $v$. 
The optimal value for $c$ is $\sqrt{n}$ and can be found by simple analysis. 
Summing over all vertices, yields $\mathcal{O}(m+n^{3/2})$ overall time per iteration.

If we use the graph that is $h$ levels beneath the current level for approximating $r$ values, we need $\mathcal{O}(d(v) + c + n/c^h)$ time to compute the new coordinate for a vertex $v$. 
Here we assume again that each cluster in the hierarchy contains $c$ vertices.
As before, a simple analysis yields that the optimal value for $c$ is $n^{\frac{1}{h+1}}$. 
Summing over all vertices, yields $\mathcal{O}(m+  n^{\frac{h+2}{h+1}})$ overall time.
\section{Basic Properties of the Benchmark Set}
\begin{table}[h]
\footnotesize
\centering
\caption{Basic properties of the benchmark set with a rough type classification.}
\label{tab:graphstable}
\begin{tabular}{|l|r|r||r||r|}
\hline
graph & $n$ & $m$ & Type & Ref. \\
\hline
\hline
\multicolumn{5}{|c|}{Small Graphs} \\
\hline
btree                                  & \numprint{1023}    & \numprint{1022}    & Binary Tree            & \cite{UFsparsematrixcollection}\\
1138\_bus                              & \numprint{1138}    & \numprint{1358}    & Power System           & \cite{UFsparsematrixcollection}\\
USpowerGrid                            & \numprint{4941}    & \numprint{6594}    & US Power Grid          & \cite{UFsparsematrixcollection}\\
3elt                                   & \numprint{4960}    & \numprint{13722}   & Airfoil                & \cite{soper2004combined}\\
commanche                              & \numprint{7920}    & \numprint{11880}   & Helicopter             & \cite{UFsparsematrixcollection}\\
\hline
\multicolumn{5}{|c|}{Medium Graphs} \\
\hline
bcsstk31                               & \numprint{35586}   & \numprint{572913}  & Automobile Component   & \cite{soper2004combined}\\
fe\_pwt                                & \numprint{36519}   & \numprint{144794}  & Structural Problem     & \cite{soper2004combined}\\
del16                                  & \numprint{65536}   & \numprint{196575}  & Delaunay Triangulation & \cite{kappa} \\
luxembourg                             & \numprint{114599}  & \numprint{119666}  & Road Network           & \cite{benchmarksfornetworksanalysis}\\
\hline
\multicolumn{5}{|c|}{Large Graphs} \\
\hline
nyc                                    & \numprint{264346}  & \numprint{365050}  & Road Network           & \cite{demetrescu2009shortest}\\
auto                                   & \numprint{448695}  & \numprint{3314611} & Automobile             & \cite{soper2004combined}\\
del20                                  & \numprint{1048576} & \numprint{3145686} & Delaunay Triangulation & \cite{kappa} \\
\hline
\end{tabular}
\end{table}

\vfill
\pagebreak

\section{Detailed Experimental Results}
\begin{table}[h!]
\footnotesize
\centering
\scalebox{0.8}{
\begin{tabular}{|r|l|r|r|r|r|r|r||}

  \hline
  $h$ & graph & $F(x)$ & $M(x)$& $t$[s]\\
  \hline
  \hline

0      & bcsstk31   & \numprint{31507}K  & \numprint{-14925}K  & \numprint{5.69} \\
1      & bcsstk31   & \numprint{31547}K  & \numprint{-14921}K  & \numprint{4.16} \\
2      & bcsstk31   & \numprint{31887}K  & \numprint{-14931}K  & \numprint{2.77} \\
3      & bcsstk31   & \numprint{30938}K  & \numprint{-14935}K  & \numprint{2.28} \\
4      & bcsstk31   & \numprint{31449}K  & \numprint{-14933}K  & \numprint{2.01} \\
5      & bcsstk31   & \numprint{31819}K  & \numprint{-14921}K  & \numprint{1.88} \\
6      & bcsstk31   & \numprint{31894}K  & \numprint{-14919}K  & \numprint{1.81} \\
7      & bcsstk31   & \numprint{32156}K  & \numprint{-14912}K  & \numprint{1.82} \\
8      & bcsstk31   & \numprint{33574}K  & \numprint{-14888}K  & \numprint{1.88} \\
9      & bcsstk31   & \numprint{34306}K  & \numprint{-14877}K  & \numprint{1.86} \\
10     & bcsstk31   & \numprint{35316}K  & \numprint{-14861}K  & \numprint{1.86} \\
11     & bcsstk31   & \numprint{36086}K  & \numprint{-14850}K  & \numprint{2.09} \\
\hline
\hline
0      & fe\_pwt    & \numprint{51501}K  & \numprint{-23850}K  & \numprint{4.44} \\
1      & fe\_pwt    & \numprint{50329}K  & \numprint{-23867}K  & \numprint{2.64} \\
2      & fe\_pwt    & \numprint{50926}K  & \numprint{-23855}K  & \numprint{1.67} \\
3      & fe\_pwt    & \numprint{50317}K  & \numprint{-23864}K  & \numprint{1.11} \\
4      & fe\_pwt    & \numprint{49620}K  & \numprint{-23874}K  & \numprint{0.88} \\
5      & fe\_pwt    & \numprint{50420}K  & \numprint{-23861}K  & \numprint{0.77} \\
6      & fe\_pwt    & \numprint{50927}K  & \numprint{-23852}K  & \numprint{0.69} \\
7      & fe\_pwt    & \numprint{50885}K  & \numprint{-23850}K  & \numprint{0.59} \\
8      & fe\_pwt    & \numprint{49889}K  & \numprint{-23862}K  & \numprint{0.68} \\
9      & fe\_pwt    & \numprint{50174}K  & \numprint{-23857}K  & \numprint{0.63} \\
10     & fe\_pwt    & \numprint{49464}K  & \numprint{-23867}K  & \numprint{0.74} \\
11     & fe\_pwt    & \numprint{48727}K  & \numprint{-23879}K  & \numprint{0.80} \\
\hline
\hline
0      & del16      & \numprint{716997}K & \numprint{-58137}K  & \numprint{13.76} \\
1      & del16      & \numprint{716391}K & \numprint{-58146}K  & \numprint{7.71} \\
2      & del16      & \numprint{724073}K & \numprint{-58031}K  & \numprint{4.40} \\
3      & del16      & \numprint{727549}K & \numprint{-57980}K  & \numprint{2.69} \\
4      & del16      & \numprint{728283}K & \numprint{-57968}K  & \numprint{1.95} \\
5      & del16      & \numprint{729923}K & \numprint{-57940}K  & \numprint{1.47} \\
6      & del16      & \numprint{733832}K & \numprint{-57884}K  & \numprint{1.14} \\
7      & del16      & \numprint{729503}K & \numprint{-57948}K  & \numprint{1.17} \\
8      & del16      & \numprint{725548}K & \numprint{-58005}K  & \numprint{0.96} \\
9      & del16      & \numprint{722181}K & \numprint{-58052}K  & \numprint{0.99} \\
10     & del16      & \numprint{718768}K & \numprint{-58097}K  & \numprint{1.20} \\
11     & del16      & \numprint{715560}K & \numprint{-58138}K  & \numprint{1.41} \\
\hline
\hline
0      & luxembourg & \numprint{503465}K & \numprint{-310576}K & \numprint{41.02} \\
1      & luxembourg & \numprint{502360}K & \numprint{-310585}K & \numprint{22.86} \\
2      & luxembourg & \numprint{501632}K & \numprint{-310615}K & \numprint{12.64} \\
3      & luxembourg & \numprint{502340}K & \numprint{-310596}K & \numprint{6.85} \\
4      & luxembourg & \numprint{498744}K & \numprint{-310659}K & \numprint{4.04} \\
5      & luxembourg & \numprint{505175}K & \numprint{-310546}K & \numprint{2.46} \\
6      & luxembourg & \numprint{518992}K & \numprint{-310370}K & \numprint{1.73} \\
7      & luxembourg & \numprint{521771}K & \numprint{-310331}K & \numprint{1.35} \\
8      & luxembourg & \numprint{528333}K & \numprint{-310215}K & \numprint{1.24} \\
9      & luxembourg & \numprint{534015}K & \numprint{-310148}K & \numprint{1.25} \\
10     & luxembourg & \numprint{537286}K & \numprint{-310106}K & \numprint{1.48} \\
11     & luxembourg & \numprint{539427}K & \numprint{-310072}K & \numprint{2.01} \\
\hline
\end{tabular}
}
\vspace*{.25cm}
\caption{Influence of parameter $h$. Smaller values are better. $F$ measures full stress, $M$ measures maxent-stress. Values marked with a K are given in thousands.}
\label{tab:influenceh}
\end{table}
{
\begin{table}
\centering
\hspace*{-.25cm}
\begin{tabular}{|l||r||r|r|r|r|r|r|}
\hline
graph & PMDS & GHN & \algorithmname$_0$ &  \algorithmname$_1$  &  \algorithmname$_2$&  \algorithmname$_7$&  \algorithmname$_{10}$   \\
\hline
\hline
btree       & \numprint{-7231}      & \numprint{-9688}      & \numprint{-9128}      & \numprint{-9086}      & \numprint{-9102}      & \numprint{-9110}      & \numprint{-9134} \\
1138\_bus   & \numprint{-10973}     & \numprint{-11917}     & \numprint{-11312}     & \numprint{-11280}     & \numprint{-11223}     & \numprint{-11236}     & \numprint{-11226} \\
USpowerG    & \numprint{-260708}    & \numprint{-265352}    & \numprint{-262664}    & \numprint{-262597}    & \numprint{-262411}    & \numprint{-262067}    & \numprint{-260938} \\
3elt        & \numprint{-276808}    & \numprint{-280535}    & \numprint{-280004}    & \numprint{-280154}    & \numprint{-280169}    & \numprint{-279721}    & \numprint{-276893} \\
commanche   & \numprint{-950570}    & \numprint{-954624}    & \numprint{-962521}    & \numprint{-962851}    & \numprint{-963107}    & \numprint{-956300}    & \numprint{-956237} \\
\hline
bcsstk31    & \numprint{-15}M  & \numprint{-15}M  & \numprint{-15}M  & \numprint{-15}M  & \numprint{-15}M  & \numprint{-15}M  & \numprint{-15}M \\
fe\_pwt     & \numprint{-24}M  & \numprint{-24}M  & \numprint{-24}M  & \numprint{-24}M  & \numprint{-24}M  & \numprint{-24}M  & \numprint{-24}M \\
del16       & \numprint{-64}M  & \numprint{-61}M  & \numprint{-58}M  & \numprint{-58}M  & \numprint{-58}M  & \numprint{-58}M  & \numprint{-58}M \\
luxembourg  & \numprint{-313}M & \numprint{-314}M & \numprint{-310}M & \numprint{-310}M & \numprint{-311}M & \numprint{-310}M & \numprint{-310}M \\
\hline
\hline
\hline

btree       & \numprint{136070}    & \numprint{63721}     & \numprint{86285}     & \numprint{87099}     & \numprint{87478}     & \numprint{85916}     & \numprint{85786} \\
1138\_bus   & \numprint{77834}     & \numprint{44822}     & \numprint{60364}     & \numprint{60631}     & \numprint{64323}     & \numprint{62337}     & \numprint{62786} \\
USpowerG    & \numprint{1123582}   & \numprint{1016164}   & \numprint{1065936}   & \numprint{1071829}   & \numprint{1073659}   & \numprint{1098712}   & \numprint{1155798} \\
3elt        & \numprint{636240}    & \numprint{581770}    & \numprint{580001}    & \numprint{568919}    & \numprint{562558}    & \numprint{580314}    & \numprint{752865} \\
commanche   & \numprint{1935570}   & \numprint{2051361}   & \numprint{1483059}   & \numprint{1460357}   & \numprint{1446978}   & \numprint{1830875}   & \numprint{1834976} \\
\hline
bcsstk31    & \numprint{33}M  & \numprint{31}M  & \numprint{32}M  & \numprint{32}M  & \numprint{32}M  & \numprint{32}M  & \numprint{35}M \\
fe\_pwt     & \numprint{24}M  & \numprint{22}M  & \numprint{52}M  & \numprint{50}M  & \numprint{51}M  & \numprint{51}M  & \numprint{49}M \\
del16       & \numprint{27}M & \numprint{50}M & \numprint{72}M & \numprint{72}M & \numprint{72}M & \numprint{73}M & \numprint{72}M \\
luxembourg  & \numprint{28}M & \numprint{23}M & \numprint{50}M & \numprint{50}M & \numprint{50}M & \numprint{52}M & \numprint{54}M \\

\hline
\end{tabular}
\hspace*{.25cm}
\smallskip
\caption{Maxent stress $M(x)$ (top) and full stress $F(x)$ (bottom) for small and medium sized graphs. Smaller is better. Values marked with an $M$ are in millions. PMDS and GHN use one core/thread, MulMent$_*$ use 32 cores (64 threads). Values with an marked with an M are shown in million.
Recall that GHN and \algoname\ optimize for $M(x)$.
}
\label{tab:compmaxentstress}
\end{table}
}
\begin{table}
\centering
\begin{tabular}{|l||r||r|r|r|r|r|r|r|}
\hline
graph & PMDS & GHN & \algorithmname$_0$ &  \algorithmname$_1$  &  \algorithmname$_2$&  \algorithmname$_7$&  \algorithmname$_{10}$   \\
\hline
\hline
btree      & 0.02  & 1.14    & 0.10    & 0.17    & 0.11   & 0.11  & 0.11\\
1138\_bus  & 0.04  & 1.41    & 0.15    & 0.13    & 0.10   & 0.11  & 0.11\\
USpowerG   & 0.14  & 3.82    & 0.29    & 0.23    & 0.18   & 0.15  & 0.18\\
3elt       & 0.16  & 3.45    & 0.21    & 0.19    & 0.16   & 0.17  & 0.17\\
commanche  & 0.24  & 5.42    & 0.32    & 0.27    & 0.18   & 0.15  & 0.18\\
\hline
\hline
bcsstk31   & 3.44  & 48.48   & 5.63    & 3.97    & 2.70   & 1.75  & 1.82\\
fe\_pwt    & 1.49  & 31.60   & 4.46    & 2.67    & 1.62   & 0.57  & 0.66\\
del16      & 2.86  & 61.42   & 13.63   & 7.75    & 4.38   & 1.01  & 1.10\\
luxembourg & 3.10  & 96.10   & 40.94   & 22.87   & 12.53  & 1.31  & 1.39\\
\hline
\hline
nyc        & 9.03  & 233.94  & 216.27  & 119.33  & 64.19  & 4.68  & 3.70\\
auto       & 41.80 & 665.67  & 613.51  & 329.08  & 179.24 & 23.64 & 20.70\\
del20      & 53.80 & 1125.03 & 3303.82 & 1749.77 & 922.10 & 51.35 & 27.01\\
\hline
\end{tabular}
\hspace*{.25cm}
\smallskip
\caption{Running times in seconds per graph. Smaller is better. PivotMDS and GHN use one thread (sequential
codes), the MulMent$_*$ algorithms use 32 cores (64 threads). Running times of GHN are without the time of PMDS (which yields input coordinates to GHN).}
\label{tab:comprunningtime}
\end{table}
\begin{table}
\centering
\begin{tabular}{|l||r||r|r|r|}
\hline
graph      & PMDS  & GHN     &  \algorithmname$_7$ & \algorithmname$_{10}$   \\
\hline
\hline
btree      & 0.02  & 1.14    & 0.07    & 0.12\\
1138\_bus  & 0.04  & 1.41    & 0.10    & 0.21\\
USpowerG   & 0.14  & 3.82    & 0.30    & 1.20\\
3elt       & 0.16  & 3.45    & 0.18    & 0.70\\
commanche  & 0.24  & 5.42    & 0.49    & 1.16\\
\hline                                           
\hline                                           
bcsstk31   & 3.44  & 48.48   & 4.41    & 7.79\\
fe\_pwt    & 1.49  & 31.60   & 2.62    & 5.92\\
del16      & 2.86  & 61.42   & 6.35    & 10.95\\
luxembourg & 3.10  & 96.10   & 16.51   & 22.32\\
\hline                                           
\hline                                           
nyc        & 9.03  & 233.94  & 78.34   & 53.24\\
auto       & 41.80 & 665.67  & 207.37  & 118.65\\
del20      & 53.80 & 1125.03 & 1101.82 & 310.29\\
\hline
\end{tabular}
\hspace*{.25cm}
\smallskip
\caption{Running times in seconds per graph. Smaller is better. All of the algorithms use one core. Running times of GHN are without the time of PMDS (which yields input coordinates to GHN).}
\label{tab:comprunningtimesinglethread}
\end{table}

\begin{table}[h]
\centering
\scalebox{0.8}{
        \hspace*{-.75cm}
\begin{tabular}{|l|l|l|l||r|r|r||r|r|r||r|r|r|r|r|r|r|r|r|r|r|r||}
\hline
graph      & $h$ & $x$ & $\mathcal{D}$ & $F_G(x)$           & $F_Q(x)$ dyn       & $F_Q(x)$ scratch   &             $M_G(x)$        &    $M_Q(x)$ dyn                 &     $M_Q(x)$ scratch                & $t_G$ & $t_{\text{dyn}}$ & $t_{\text{scratch}}$ \\
\hline
\hline
bcsstk31   & 0   & 1   & 2             & \numprint{30731}K  & \numprint{31514}K  & \numprint{32867}K  & \numprint{-14932}K  & \numprint{-13914}K  & \numprint{-13896}K  & 5.87  & 1.64  & 5.73\\
bcsstk31   & 0   & 1   & 16            & \numprint{30731}K  & \numprint{74746}K  & \numprint{71772}K  & \numprint{-14932}K  & \numprint{-7878}K   & \numprint{-7975}K   & 5.94  & 3.14  & 10.18\\
bcsstk31   & 0   & 5   & 2             & \numprint{30731}K  & \numprint{30680}K  & \numprint{32271}K  & \numprint{-14932}K  & \numprint{-13123}K  & \numprint{-13098}K  & 5.93  & 2.43  & 6.62\\
bcsstk31   & 0   & 5   & 16            & \numprint{30731}K  & \numprint{77916}K  & \numprint{75695}K  & \numprint{-14932}K  & \numprint{-6559}K   & \numprint{-6649}K   & 5.92  & 3.25  & 14.28\\
\hline
bcsstk31   & 7   & 1   & 2             & \numprint{32421}K  & \numprint{33517}K  & \numprint{32888}K  & \numprint{-14903}K  & \numprint{-13882}K  & \numprint{-13894}K  & 1.86  & 1.51  & 1.72\\
bcsstk31   & 7   & 1   & 16            & \numprint{32421}K  & \numprint{74908}K  & \numprint{71692}K  & \numprint{-14903}K  & \numprint{-7869}K   & \numprint{-7975}K   & 1.87  & 1.61  & 1.97\\
bcsstk31   & 7   & 5   & 2             & \numprint{32421}K  & \numprint{32638}K  & \numprint{31702}K  & \numprint{-14903}K  & \numprint{-13092}K  & \numprint{-13098}K  & 1.87  & 1.69  & 1.90\\
bcsstk31   & 7   & 5   & 16            & \numprint{32421}K  & \numprint{78223}K  & \numprint{78906}K  & \numprint{-14903}K  & \numprint{-6543}K   & \numprint{-6565}K   & 1.96  & 2.10  & 2.94\\
\hline
\hline
del16      & 0   & 1   & 2             & \numprint{668857}K & \numprint{664866}K & \numprint{737076}K & \numprint{-58842}K  & \numprint{-58687}K  & \numprint{-57623}K  & 14.31 & 2.81  & 14.04\\
del16      & 0   & 1   & 16            & \numprint{668857}K & \numprint{492192}K & \numprint{250399}K & \numprint{-58842}K  & \numprint{-48309}K  & \numprint{-51443}K  & 14.21 & 5.27  & 14.45\\
del16      & 0   & 5   & 2             & \numprint{668857}K & \numprint{662893}K & \numprint{701075}K & \numprint{-58842}K  & \numprint{-57787}K  & \numprint{-57234}K  & 14.28 & 2.82  & 14.07\\
del16      & 0   & 5   & 16            & \numprint{668857}K & \numprint{458453}K & \numprint{250483}K & \numprint{-58842}K  & \numprint{-40845}K  & \numprint{-43516}K  & 14.30 & 7.73  & 18.88\\
\hline
del16      & 7   & 1   & 2             & \numprint{657691}K & \numprint{653806}K & \numprint{728689}K & \numprint{-58998}K  & \numprint{-58841}K  & \numprint{-57746}K  & 1.24  & 0.68  & 1.02\\
del16      & 7   & 1   & 16            & \numprint{657691}K & \numprint{484715}K & \numprint{250234}K & \numprint{-58998}K  & \numprint{-48389}K  & \numprint{-51455}K  & 1.24  & 0.74  & 1.13\\
del16      & 7   & 5   & 2             & \numprint{657691}K & \numprint{651547}K & \numprint{715301}K & \numprint{-58998}K  & \numprint{-57946}K  & \numprint{-57015}K  & 1.16  & 0.69  & 1.04\\
del16      & 7   & 5   & 16            & \numprint{657691}K & \numprint{451676}K & \numprint{242099}K & \numprint{-58998}K  & \numprint{-40918}K  & \numprint{-43650}K  & 1.27  & 0.88  & 1.28\\
\hline
\hline
fe\_pwt    & 0   & 1   & 2             & \numprint{51015}K  & \numprint{42517}K  & \numprint{40472}K  & \numprint{-23858}K  & \numprint{-23283}K  & \numprint{-23305}K  & 4.67  & 1.00  & 4.57\\
fe\_pwt    & 0   & 1   & 16            & \numprint{51015}K  & \numprint{30982}K  & \numprint{27235}K  & \numprint{-23858}K  & \numprint{-17935}K  & \numprint{-17985}K  & 4.67  & 1.01  & 4.59\\
fe\_pwt    & 0   & 5   & 2             & \numprint{51015}K  & \numprint{37268}K  & \numprint{38068}K  & \numprint{-23858}K  & \numprint{-22543}K  & \numprint{-22530}K  & 4.72  & 1.01  & 4.59\\
fe\_pwt    & 0   & 5   & 16            & \numprint{51015}K  & \numprint{33316}K  & \numprint{29371}K  & \numprint{-23858}K  & \numprint{-15423}K  & \numprint{-15477}K  & 4.73  & 1.78  & 4.68\\
\hline
fe\_pwt    & 7   & 1   & 2             & \numprint{49563}K  & \numprint{41148}K  & \numprint{38419}K  & \numprint{-23870}K  & \numprint{-23295}K  & \numprint{-23328}K  & 0.67  & 0.41  & 0.58\\
fe\_pwt    & 7   & 1   & 16            & \numprint{49563}K  & \numprint{30930}K  & \numprint{27818}K  & \numprint{-23870}K  & \numprint{-17932}K  & \numprint{-17974}K  & 0.76  & 0.42  & 0.63\\
fe\_pwt    & 7   & 5   & 2             & \numprint{49563}K  & \numprint{36242}K  & \numprint{36014}K  & \numprint{-23870}K  & \numprint{-22551}K  & \numprint{-22551}K  & 0.72  & 0.43  & 0.64\\
fe\_pwt    & 7   & 5   & 16            & \numprint{49563}K  & \numprint{32843}K  & \numprint{30130}K  & \numprint{-23870}K  & \numprint{-15426}K  & \numprint{-15473}K  & 0.76  & 0.52  & 0.68\\
\hline
\hline
luxembourg & 0   & 1   & 2             & \numprint{511117}K & \numprint{522212}K & \numprint{526761}K & \numprint{-310488}K & \numprint{-311406}K & \numprint{-311292}K & 42.60 & 7.69  & 42.31\\
luxembourg & 0   & 1   & 16            & \numprint{511117}K & \numprint{549426}K & \numprint{556099}K & \numprint{-310488}K & \numprint{-307121}K & \numprint{-307076}K & 42.50 & 7.71  & 42.49\\
luxembourg & 0   & 5   & 2             & \numprint{511117}K & \numprint{788036}K & \numprint{881315}K & \numprint{-310488}K & \numprint{-318330}K & \numprint{-317828}K & 42.60 & 7.69  & 42.39\\
luxembourg & 0   & 5   & 16            & \numprint{511117}K & \numprint{619716}K & \numprint{551834}K & \numprint{-310488}K & \numprint{-297447}K & \numprint{-298532}K & 42.49 & 7.71  & 42.57\\
\hline
luxembourg & 7   & 1   & 2             & \numprint{516248}K & \numprint{526999}K & \numprint{546575}K & \numprint{-310400}K & \numprint{-311324}K & \numprint{-311045}K & 1.37  & 0.55  & 1.23\\
luxembourg & 7   & 1   & 16            & \numprint{516248}K & \numprint{551360}K & \numprint{592054}K & \numprint{-310400}K & \numprint{-307071}K & \numprint{-306588}K & 1.48  & 0.56  & 1.31\\
luxembourg & 7   & 5   & 2             & \numprint{516248}K & \numprint{789014}K & \numprint{883948}K & \numprint{-310400}K & \numprint{-318323}K & \numprint{-317740}K & 1.38  & 0.53  & 1.26\\
luxembourg & 7   & 5   & 16            & \numprint{516248}K & \numprint{617187}K & \numprint{552312}K & \numprint{-310400}K & \numprint{-297457}K & \numprint{-298503}K & 1.35  & 0.58  & 1.29\\
\hline
\end{tabular}
}
\hspace*{.25cm}
\caption{Detailed per instances results for dynamic graph experiments. Smaller values are better. Values are given in thousands. $F_*$ measures full stress, $M_*$ measures maxent-stress, \emph{dyn} refers to the algorithm that updates the given coordinates, \emph{scratch} to the algorithm that discards the given coordinates and updates the layout. Running times are given in seconds. Values marked with a K are given in thousand.}
\label{tab:dynamicresults}
\end{table}
\
\vfill
\pagebreak
\
\vfill
\pagebreak
\section{Additional Drawings}
\begin{figure}
\centering
\includegraphics[width=4.5cm]{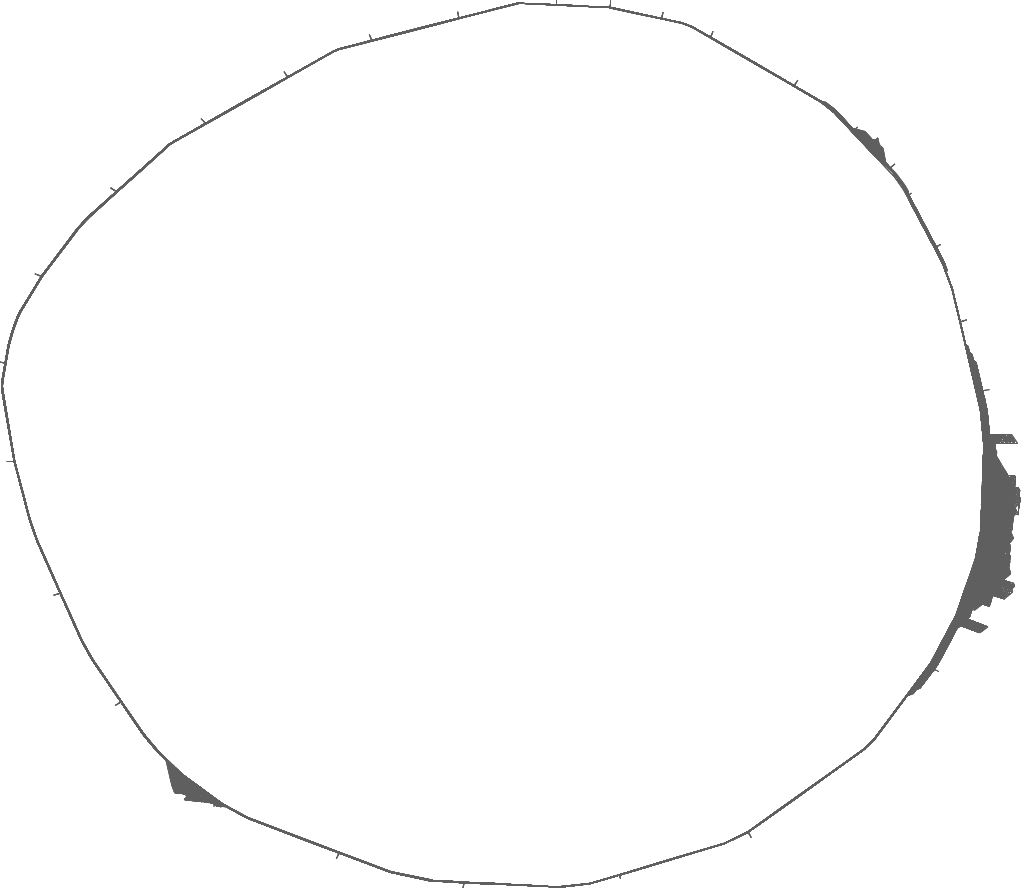}\hspace*{.75cm}\includegraphics[width=5.5cm]{drawings/pmds_bcsstk31.png}  \\
                       \vspace*{.1cm}
\includegraphics[width=4.5cm]{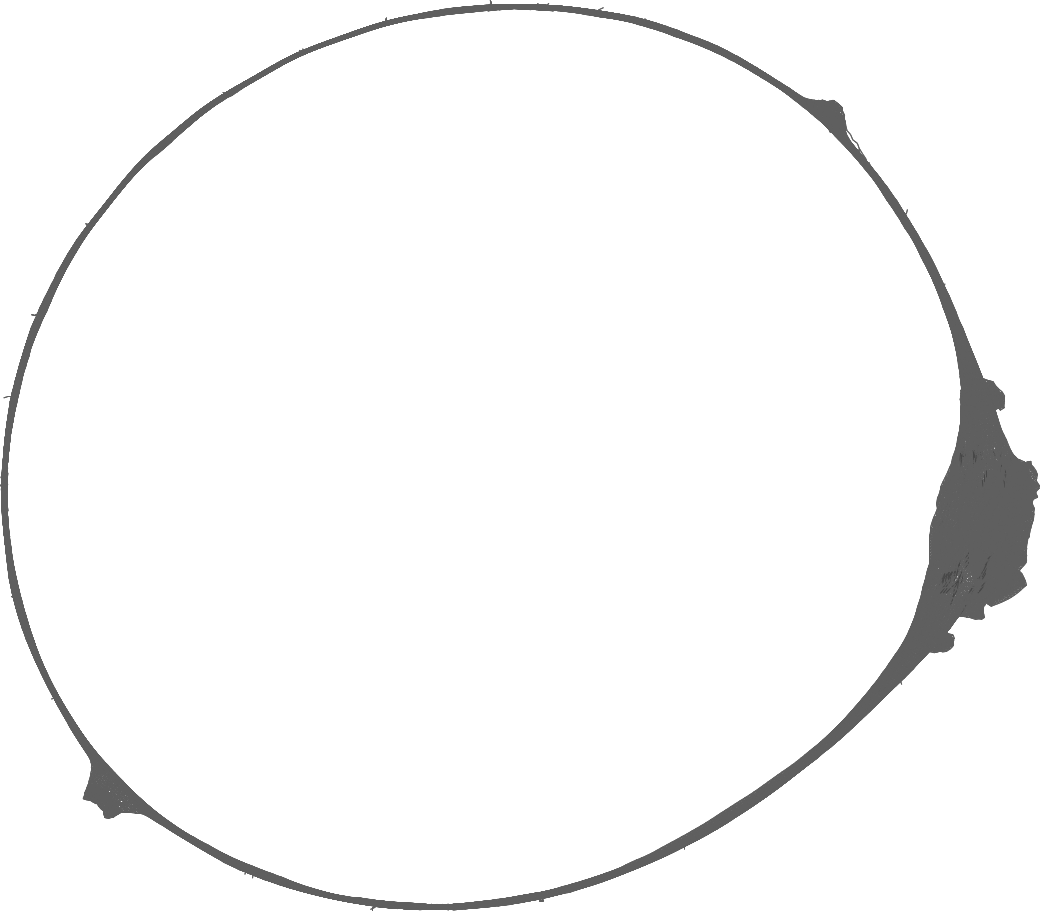}\hspace*{.75cm}\includegraphics[width=5.5cm]{drawings/maxent_bcsstk31.png} \\
                       \vspace*{.1cm}
\includegraphics[width=4.5cm]{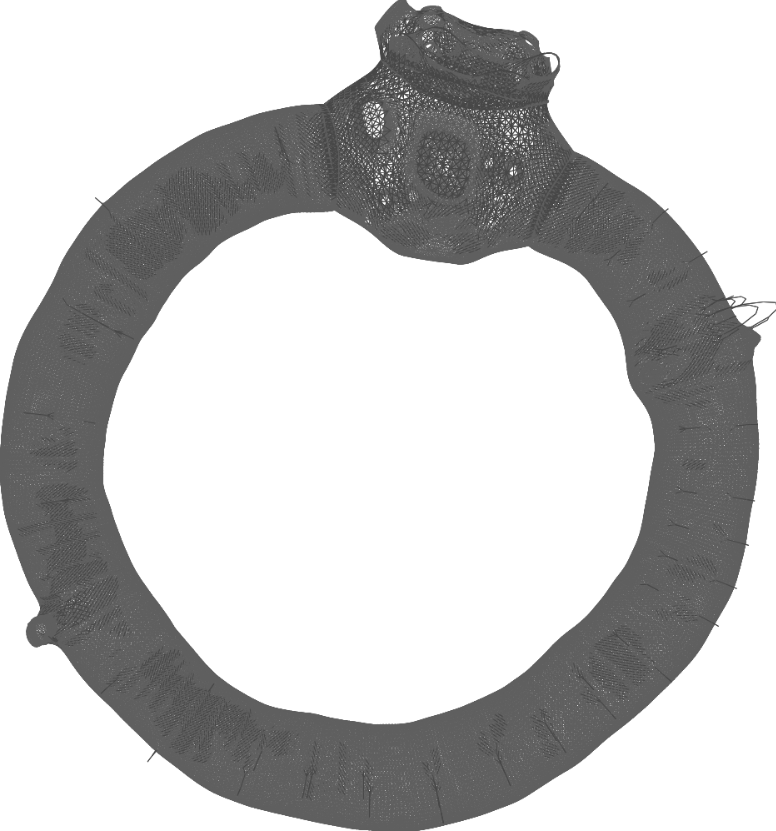}\hspace*{.75cm}\includegraphics[width=5.85cm]{drawings/mulent_bcsstk31.png}
\caption{Drawings of the largest connected components of fe\_pwt (LHS) and bcsstk31 (RHS). From top to bottom: PMDS, MaxEnt, MulMent}
\label{fig:fe_pwt}
\end{figure}
\begin{figure}
\centering
\includegraphics[width=6.5cm]{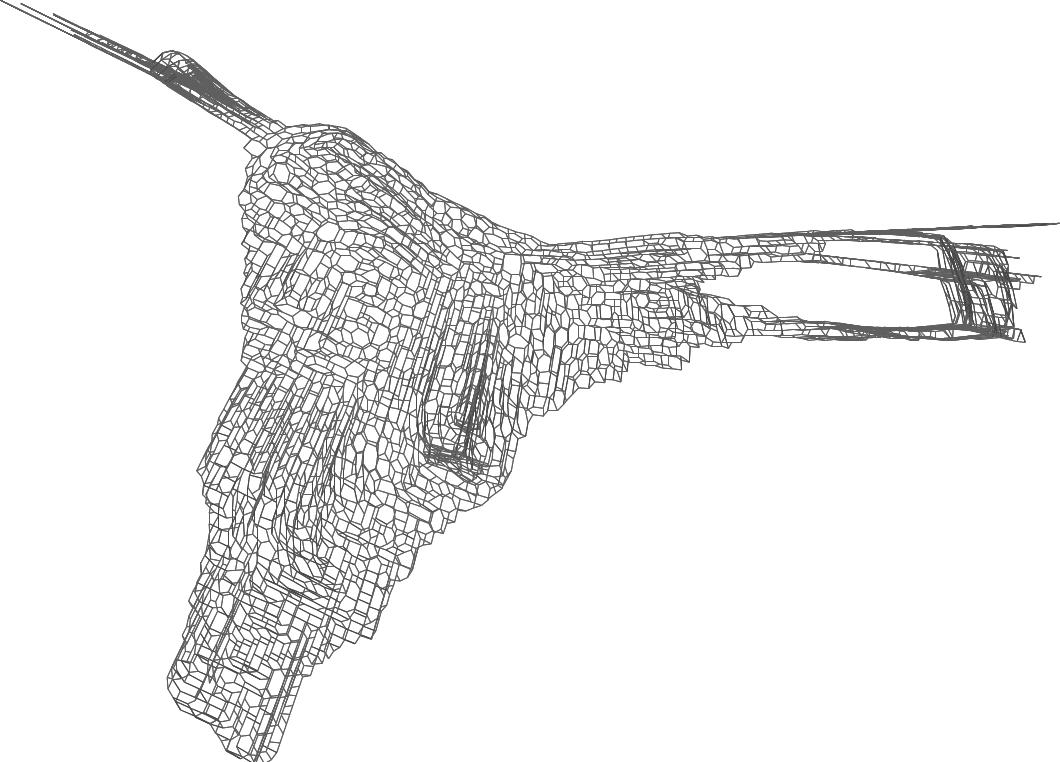}\hspace*{.75cm}\includegraphics[width=5.5cm]{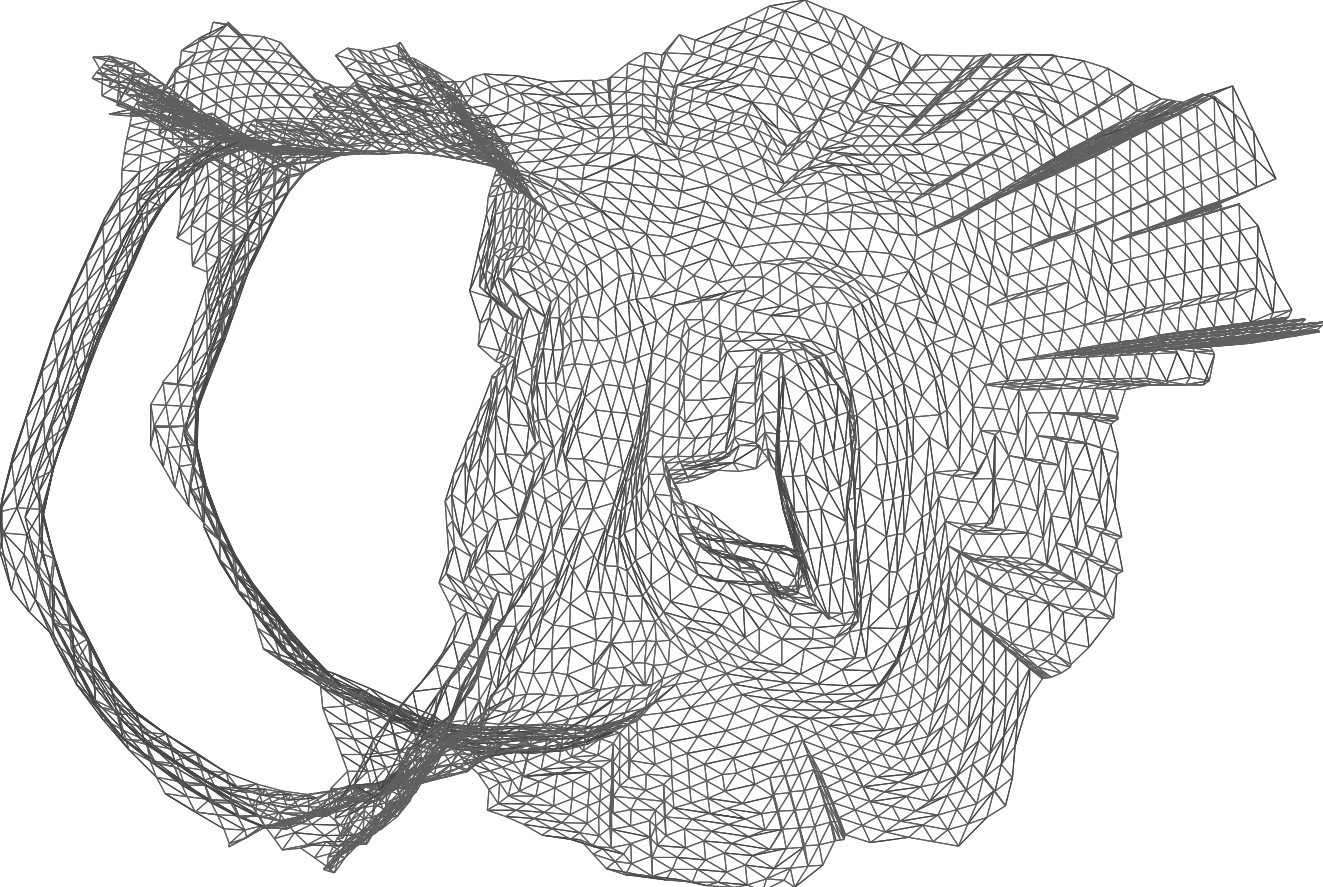}  \\
                       \vspace*{.1cm}
\includegraphics[width=6.5cm]{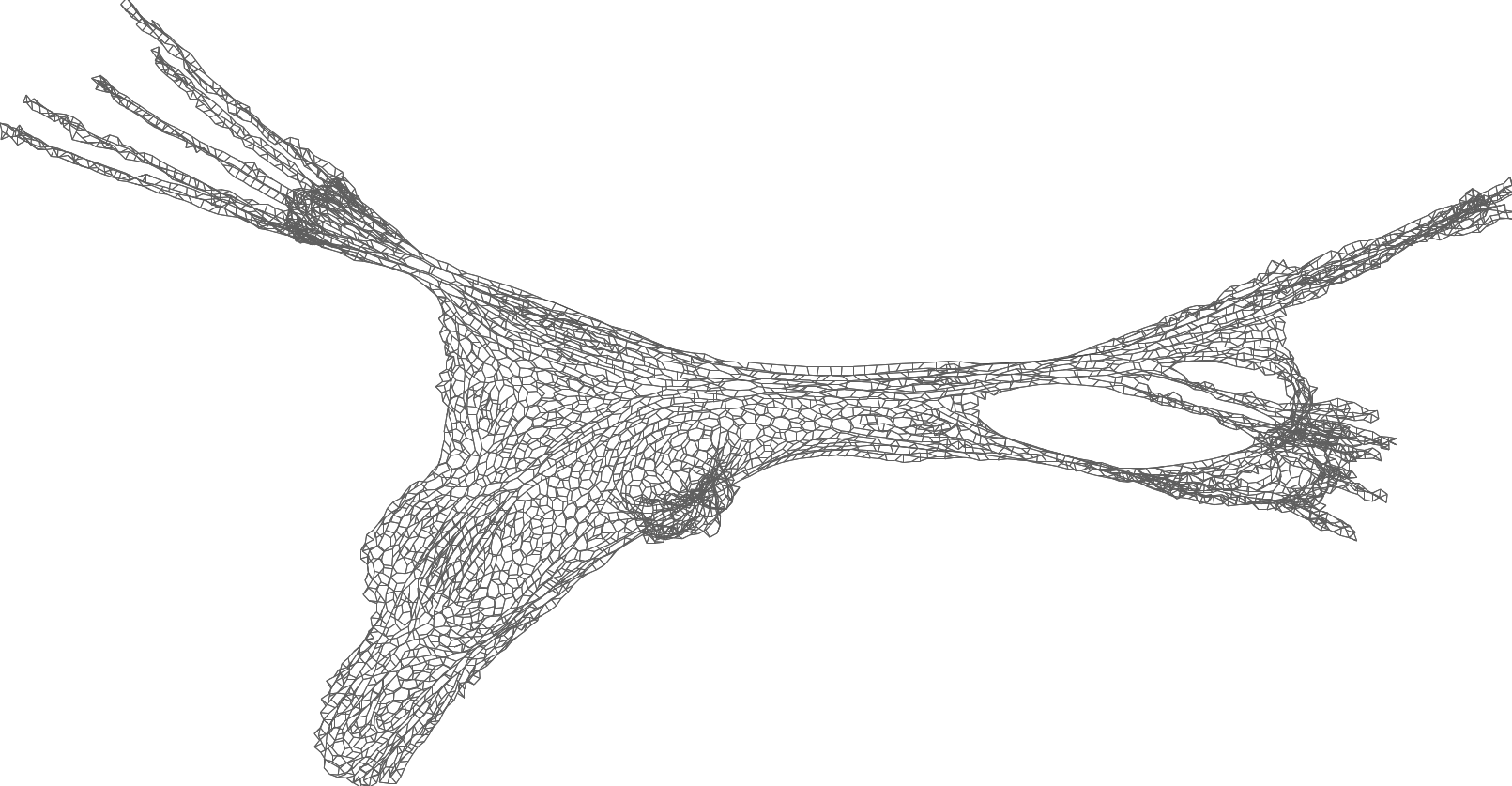}\hspace*{.75cm}\includegraphics[width=5.5cm]{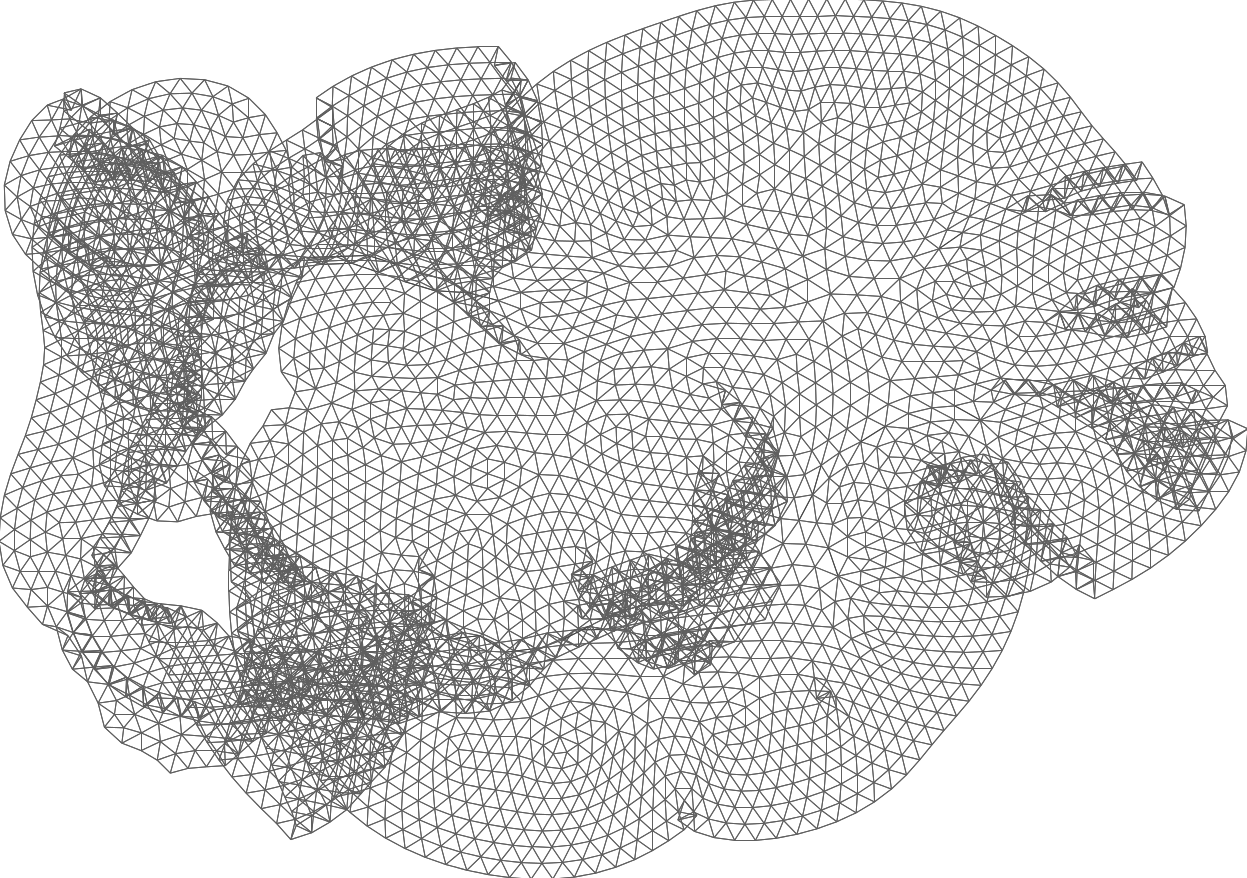} \\
                       \vspace*{.1cm}
\hspace*{1cm}\includegraphics[width=4cm]{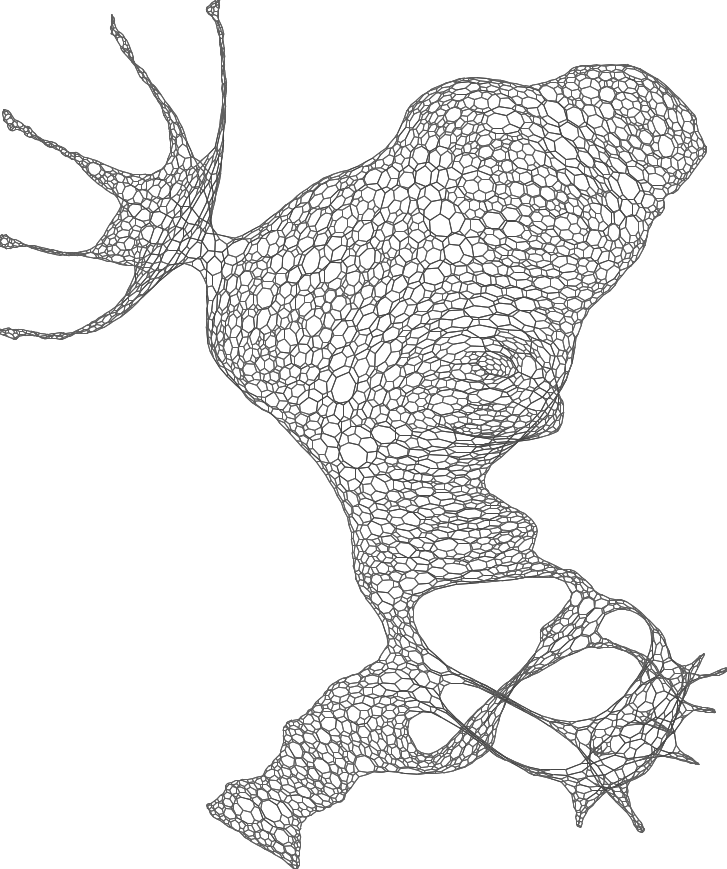}\hspace*{1.75cm}\includegraphics[width=5.85cm]{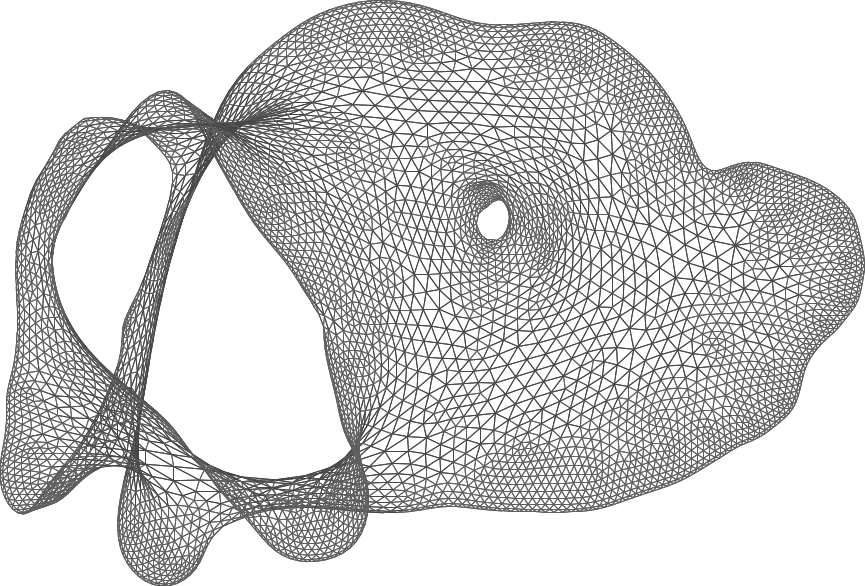}
\caption{Drawings of the largest connected components of commanche (LHS) and 3elt (RHS). From top to bottom: PMDS, MaxEnt, MulMent}
\label{fig:fe_pwt}
\end{figure}

\begin{figure}
\centering
\includegraphics[width=6cm]{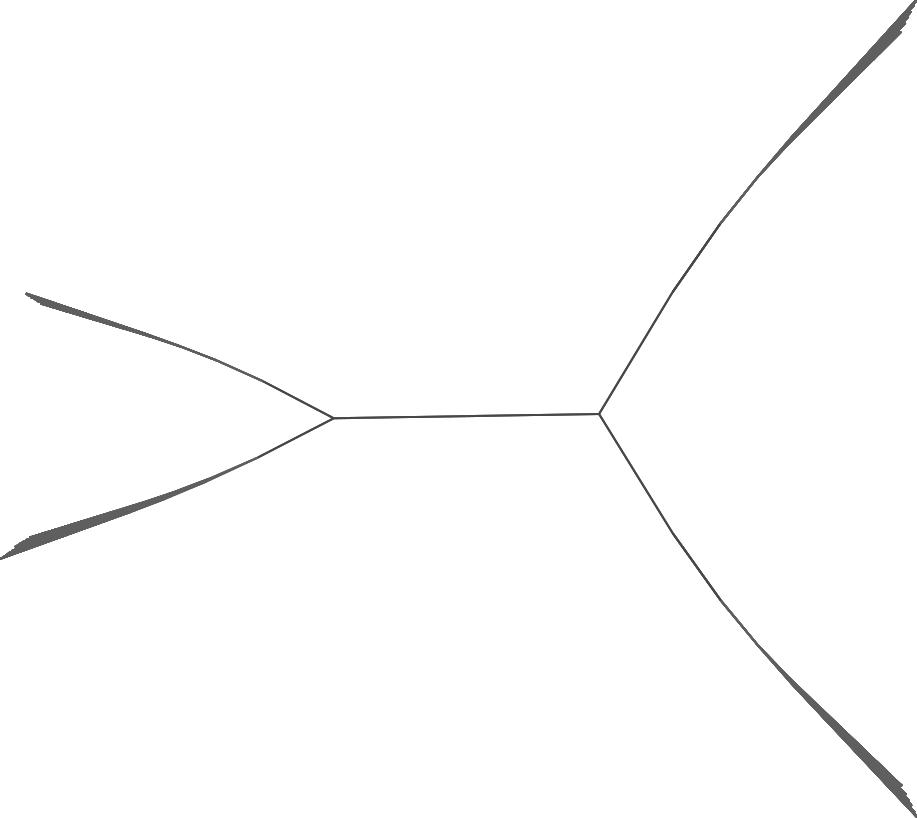}\hspace*{.75cm}\includegraphics[width=5.5cm]{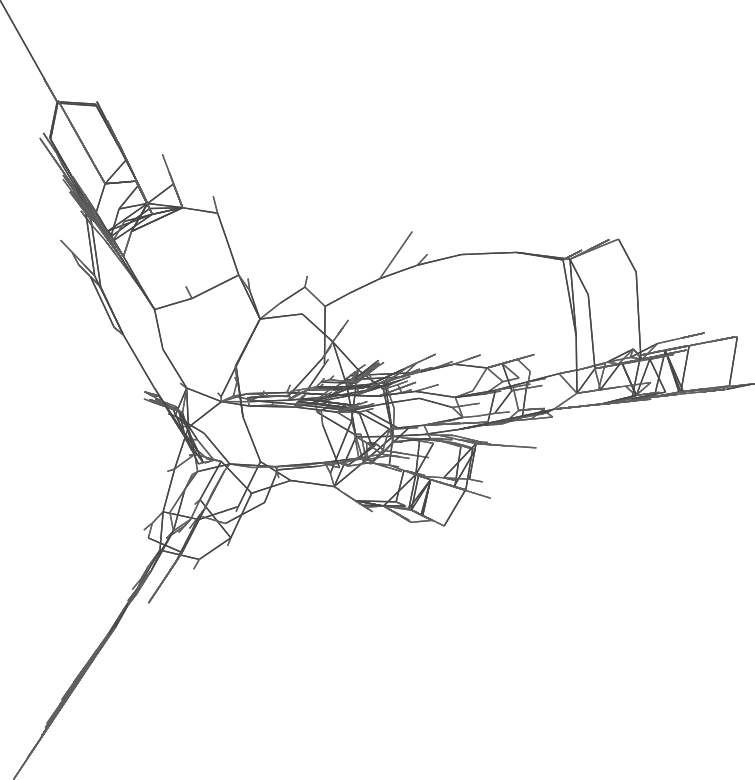}  \\
                       \vspace*{.1cm}
\includegraphics[width=6.5cm]{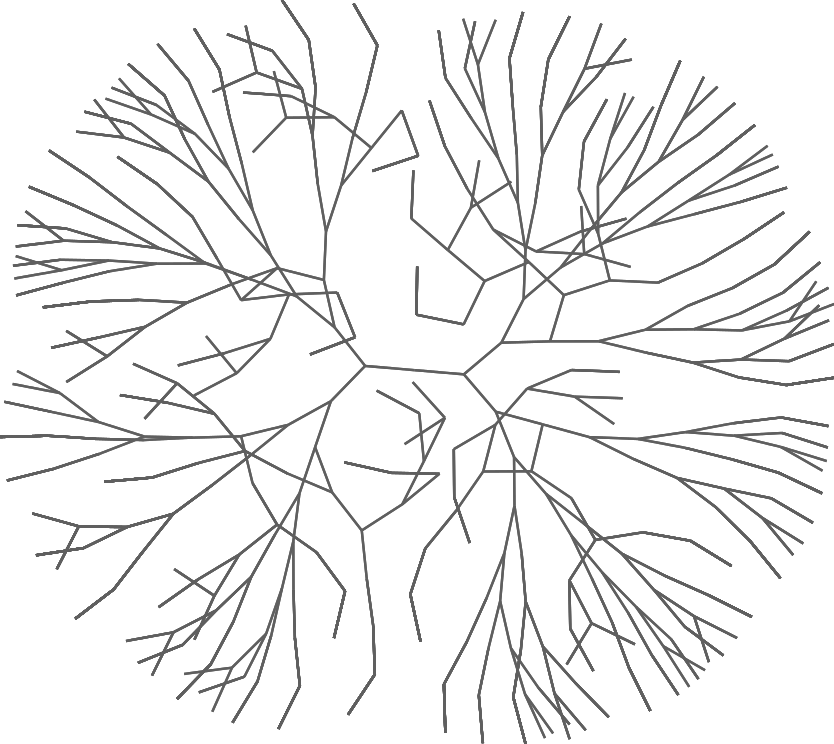}\hspace*{.75cm}\includegraphics[width=5.5cm]{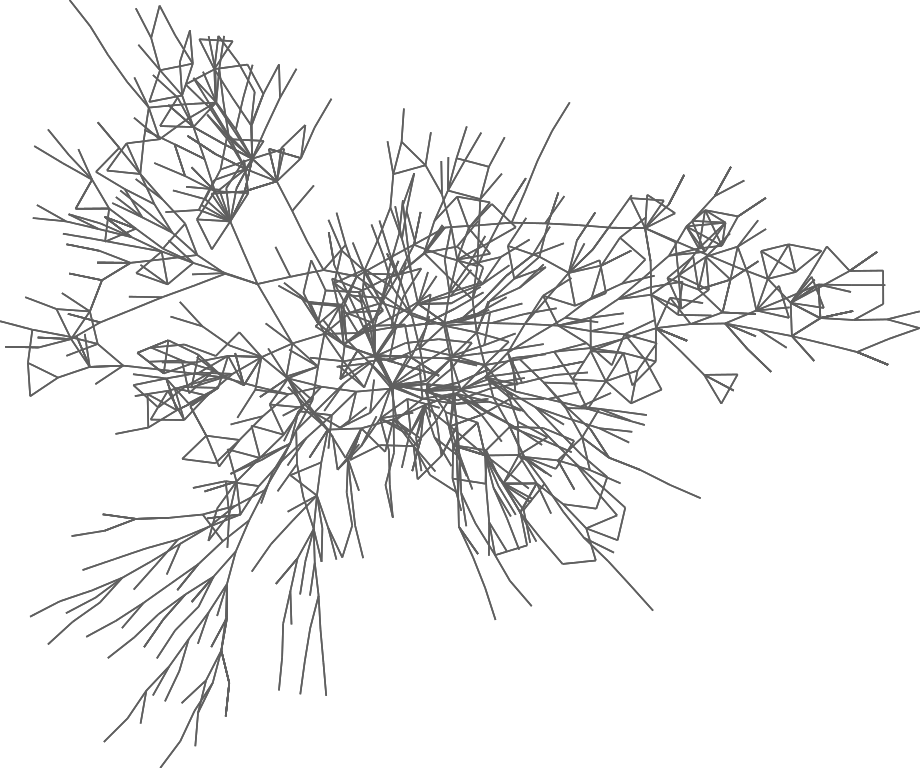} \\
                       \vspace*{.1cm}
\hspace*{1cm}\includegraphics[width=4cm]{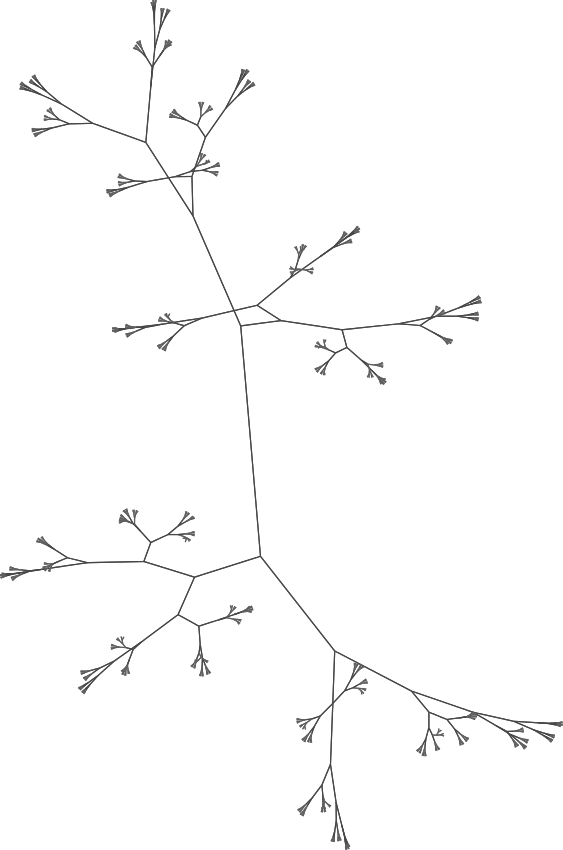}\hspace*{1.75cm}\includegraphics[width=5.85cm]{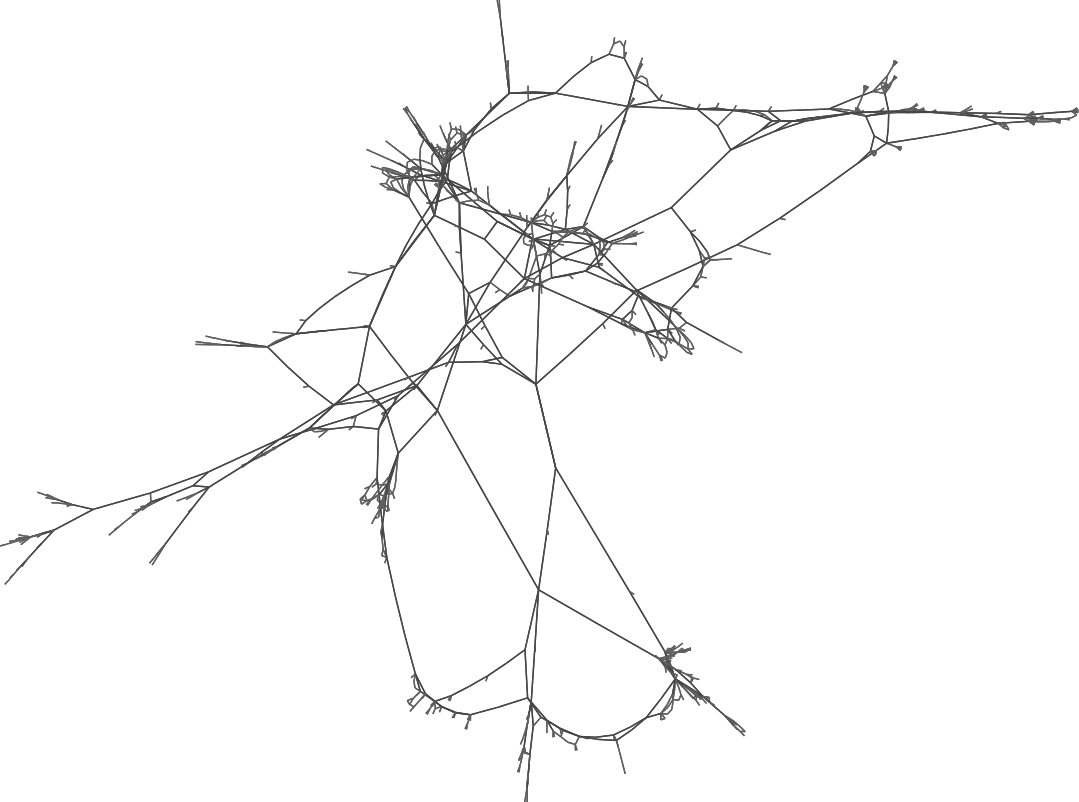}
\caption{Drawings of the largest connected components of btree (LHS) and 1138\_bus (RHS). From top to bottom: PMDS, MaxEnt, MulMent}
\label{fig:fe_pwt}
\end{figure}

\end{appendix}

\end{document}